\documentclass[10pt,twocolumn,letterpaper]{article}
\usepackage[textwidth=41pc,textheight=694.75pt,left=53.25pt,top=48pt,headheight=0pt,headsep=0pt,footskip=24pt]{geometry}
\usepackage[T1]{fontenc}
\usepackage{times}
\usepackage{amsmath,amsthm,cite,array}
\usepackage{graphicx,url}

\usepackage[linesnumbered,ruled,vlined]{algorithm2e}
\SetKwInput{KwInput}{Input}
\SetKwInput{KwOutput}{Output}
\SetVlineSkip{0pt}
\usepackage{caption}
\DeclareCaptionFont{compact}{\fontsize{8.25}{10}\selectfont}
\usepackage[hidelinks]{hyperref}

\makeatletter
\renewcommand{\normalsize}{\@setfontsize\normalsize{10.25bp}{12bp}\abovedisplayskip=5pt plus4pt minus2pt\belowdisplayskip=\abovedisplayskip\abovedisplayshortskip=0pt plus4pt\belowdisplayshortskip=5pt plus4pt minus2pt}
\renewcommand{\small}{\@setfontsize\small{8}{12pt}}
\renewcommand{\footnotesize}{\@setfontsize\footnotesize{8.25}{10pt}}
\normalsize
\renewcommand{\section}{\@startsection{section}{1}{0pt}{19.25pt plus2pt minus2pt}{5pt}{\normalfont\sffamily\fontsize{10}{12}\selectfont\raggedright}}
\renewcommand{\subsection}{\@startsection{subsection}{2}{0pt}{19.25pt plus2pt minus2pt}{5pt}{\normalfont\sffamily\fontsize{10}{12}\selectfont\raggedright}}
\renewcommand{\subsubsection}{\@startsection{subsubsection}{3}{0pt}{11pt plus2pt minus2pt}{1pt}{\normalfont\fontsize{10}{12}\selectfont\itshape\raggedright}}

\renewcommand{\@seccntformat}[1]{\csname the#1\endcsname.\hspace{0.5em}}
\renewcommand{\@maketitle}{\begin{center}{\sffamily\fontsize{18.25}{24}\selectfont\@title\par}\vspace{12pt}{\fontsize{8.25}{10}\selectfont\@author\par}\end{center}\vspace{12pt}}
\makeatother
\renewenvironment{abstract}{\par\begingroup\fontsize{8.25}{10}\selectfont\bfseries\noindent\textit{Abstract---}\ignorespaces}{\par\endgroup\vspace{1.34ex}}
\AddToHook{env/thebibliography/begin}{\fontsize{8}{10}\selectfont}
\let\originalappendix\appendix
\renewcommand{\appendix}{\originalappendix\raggedbottom\section*{Appendix}}
\newtheoremstyle{compact}{6pt}{6pt}{\normalfont}{}{\scshape}{.}{0.5em}{}
\theoremstyle{compact}
\newtheorem{definition}{Definition}
\newtheorem{insight}{Insight}

\newcommand{\mcf}{\ensuremath{{\sf mcf}}}
\newcommand{\mdf}{\ensuremath{{\sf mdf}}}
\newcommand{\general}{\ensuremath{\sf infra}}
\newcommand{\attackflow}{\ensuremath{\sf af}}
\newcommand{\Lat}{\ensuremath{{\sf L}}}
\newcommand{\AT}{\ensuremath{{\sf AT}}}
\newcommand{\at}{\ensuremath{{\sf at}}}
\newcommand{\dt}{\ensuremath{{\sf sc}}}
\newcommand{\flag}{\ensuremath{{\sf flag}}}

\renewcommand{\L}{\ensuremath{{\cal L}}}

\title{SCRM: An Actionable Framework for Space Cyber Risk Management}
\author{\textbf{Ekzhin Ear}\qquad\textbf{Caleb Chang}\qquad\textbf{Shouhuai Xu}\\[4pt]
\{eear, cchang, sxu\}@uccs.edu\\[4pt]
The Laboratory for Cybersecurity Dynamics\\
Department of Computer Science, University of Colorado Colorado Springs
}
\date{}

\begin{document}
\maketitle

\begin{abstract}
Space infrastructures play critical roles in modern society, including satellite communications (SATCOM).
Like the Internet, space infrastructures are vulnerable to cyber attacks, or {\em space cyber attacks}, highlighting the importance of managing cyber risks to space infrastructures, or {\em space cyber risks}. However, adequately managing space cyber risks is an open problem. In this paper, we propose the first \underline{S}pace \underline{C}yber \underline{R}isk \underline{M}anagement (SCRM) framework to systematically analyze and mitigate space cyber risks. The framework models space infrastructures, space missions, and space cyber attacks, while offering algorithms for space mission risk analysis and hardening. We demonstrate its usefulness by conducting a case study on
real-world space cyber attacks against the SATCOM infrastructure implemented in our testbed. Our results show, among other things, that: (i) dealing with space cyber attack cascading effects is essential to space cyber risk management; (ii) the framework can effectively harden space missions; (iii) National Institute of Standards and Technology (NIST) security controls can effectively mitigate space cyber risks.
\end{abstract}

\section{Introduction}

Cyber threats against space infrastructures and systems are real, as evidenced by many incidents (see, e.g.,~\cite{ear2023characterizing, pavur2022building, falco2021security}). This highlights the importance of designing principled solutions to the space cyber risk management problem.
Towards this goal, the first significant
effort
is
the Space Attack Research and Tactic Analysis (SPARTA) framework~\cite{sparta_homepage} developed by The Aerospace Corporation, or more specifically its space cyber Notional Risk Scores (NRS) system~\cite{sparta_notional}, which aims to associate notional (i.e., hypothetical but plausible) cyber risks to attack techniques against space systems represents the state of the art.
However, a study~\cite{Ear2024NRS,XuIEEECSR2025-NRS} conducted by academia-industry researchers including the inventor of NRS,
shows that NRS only represents a first step towards
a principled solution to the space cyber risk management problem. This motivates the present study, which aims to propose a principled space cyber risk management framework that goes much beyond NRS.

\noindent{\bf Our Contributions}.
This paper makes three  contributions.
First, we propose a novel and {\em actionable} \underline{S}pace \underline{C}yber \underline{R}isk \underline{M}anagement (SCRM) framework, where ``actionable'' means it provides detailed representations and models for real-world adoption.
The framework
has the following features.
(i) It is geared towards space infrastructures and the missions they support.
(ii) It accommodates two concepts recently introduced by other researchers \cite{XuIEEESP2025}:  {\em mission control flow}, which specifies how commands are transmitted between units of a space infrastructure, and {\em mission data flow}, which specifies how data items are processed by, and transmitted between, units of a space infrastructure, at a desired level of abstraction. (iii) It is the first study that {\em explicitly} models space cyber attack {\em cascading effects}.
(iv) It has two core algorithms, one for {\em mission risk analysis} and the other for {\em mission hardening}. Both algorithms can use our sub-routines or incorporate sub-routines desired by analysts in a plug-and-play fashion.
(v) It can be applied in the design and operation phases  of space infrastructures, assuming hardening-after-launch is feasible.
(vi) It extends beyond NRS by accommodating the {\em cascading effects} incurred by successful space cyber attacks.

Second, we demonstrate the actionability and usefulness of the framework by implementing the algorithms and applying the framework to manage space cyber risks, while leveraging our  space testbed for validation purposes.
Our results show:
(i) the framework can accommodate both ATT\&CK and SPARTA {\em attack techniques} in modeling real-world space cyber attacks;
(ii) space cyber attack cascading effects, or {\em cascading effects} for short,  impose a major challenge to adequate space cyber risk management;  (iii) NIST security controls
can be leveraged to harden space infrastructures and their missions;
(iv) the framework can effectively reduce space cyber risks to a tolerable level.

Third, we show that SCRM substantially advances the state of the art space cyber risk analysis and management solution, namely NRS, via a concrete scenario and quantitative metrics. Specifically, we show (in Section \ref{subsec:nrs-comparison}) that NRS underestimates risk because it cannot model the cascading effect, and  overestimates the investment that is necessary in order to harden missions.

\smallskip

\noindent{\bf Related Work}.
 In a wider context,
NIST has proposed two frameworks:
(i) the Cyber Security Framework (CSF)~\cite{NISTCSF2},
which offers 6 functions---identify, protect, detect, respond, recover, and govern;
and (ii) the Risk Management Framework (RMF)~\cite{joint2018nist800-37}, which implements CSF with security controls
~\cite{NIST800-53r5,NIST800-53A,NIST800-53B,joint2012nist800-30}.
However, these frameworks provide high-level guidance without considering space infrastructure specifics,
meaning that they are not immediately actionable without making additional efforts.

In the context of space cyber risk management, NIST has also published a sequence of guidance on adapting CSF to the space domain~\cite{scholl2021introduction}, to space-based positioning, navigation, and timing (PNT) services~\cite{bartock2021foundational}, to the satellite/spacecraft command and control systems in the ground segment~\cite{lightman2022satellite}, and to hybrid satellite networks
~\cite{mccarthy2023cybersecurity}.
Moreover, NASA~\cite{NASA-STD-1006A} offers security controls
to reduce cyber risks to satellite command authority (i.e., the ability to issue commands to satellites)
and PNT
~\cite{jones2022risk,NASA-SS-BPG, bailey2021cybersecurity}.
Although these efforts consider space specifics, they are still high level and up to practitioners to interpret them when turning them into actionable procedures or software tools. This can be justified by the introduction of SPARTA NRS~\cite{sparta_notional}, which is a significant first step at addressing the space cyber risk management problem while leveraging strengths of the frameworks mentioned above. That is, NRS represents the state of the art in space cyber risk analysis and management.

However, an academia-industry study~\cite{Ear2024NRS,XuIEEECSR2025-NRS} shows NRS~\cite{sparta_notional} is largely a manual process that substantially relies on subject matter experts (in 4 out of its 6 steps), meaning different experts will likely produce different outcomes. Moreover, NRS and its predecessors focus on {\em individual} space systems, rather than a space infrastructure where new challenges emerge (e.g., cascading effects incurred by space cyber attacks).
This study  makes a significant step forward beyond NRS,
by proposing an actionable framework
that can be automated in 12 out of its 13 steps, while assuming the availability of certain inputs to its algorithms and noting that the remaining step is for identifying security controls to mitigate attack techniques (automation of which is orthogonal to our focus and left to future study). We will show, via a concrete scenario and quantitative metrics, that NRS can produce misleading results when compared with SCRM (cf. Section \ref{subsec:nrs-comparison}).
Additionally, this study advances the state of the art in its mission-centric approach, by quantifying risk based on end-to-end mission control and data flows and allowing hardening decisions to be tied directly to mission success probability rather than isolated vulnerabilities.

Loosely related prior studies are the ones that focus on space cyber attacks (e.g.,~\cite{ear2023characterizing,knez2016lessons,falco2023wannafly,falco2023scrub,falco2021cubesat,boschetti2023ground,ramdass2022cybersecurity,falco2020satellites,lee2023vulnerabilities,hansen2024guarding}).
These studies typically focus on specific space cyber attacks rather than infrastructure-wide space cyber risk management, which is our focus. At a building-block level, we leverage 3 real-world space cyber attacks reported in a recent study~\cite{ear2023characterizing} for our case study purposes.

\noindent{\bf Paper Organization}.
Section~\ref{sec:framework} describes the SCRM framework.
Section~\ref{sec:caseStudy} presents a case study
to show the actionability and usefulness of SCRM.
Section~\ref{sec:limitations} discusses limitations of this study.
Section~\ref{sec:conclusion} concludes the paper.
Appendix~\ref{appendix:acronyms} summarizes the acronyms used in the paper.

\section{The SCRM Framework}
\label{sec:framework}

\noindent{\bf Terminology and Notational Conventions}.
A space {\em infrastructure} typically consists of four segments: space, link, ground, and user.
The applications supported by space infrastructures are known as {\em payloads} or {\em application-layer missions}.
Payloads are supported by infrastructure-layer services, or {\em infrastructure-layer missions}. Each application-layer mission can be specified by one or multiple {\em control flows}, which may cut across the application and infrastructure layers, and possibly one or multiple {\em data flows}, which would stay at the application layer~\cite{XuIEEESP2025}. Each infrastructure-layer mission can also have one or multiple {\em control flows} and {\em data flows}; the data flows would stay at the infrastructure layer~\cite{XuIEEESP2025}. In this paper, {\em missions} refer to both application-layer and infrastructure-layer missions, unless explicitly stated otherwise.
{\em Mission disruption} refers to the interruption or compromise, either temporary or permanent, of a mission. In this paper we consider mission disruption incurred by space cyber attacks.

Our model is based on {\em graph theory},
and {\em directed multigraphs} in particular. We use the standard terms of {\em nodes} (i.e., vertices) and {\em arcs} (i.e., directed edges).
Multigraph means there can be multiple arcs between a same pair of nodes  because these arcs reflect different communication paths, control flows, or data flows.
We also use standard set operations, such as the union of node sets or arc sets.

\begin{figure}[!htbp]
\centering{\includegraphics[width=.6\columnwidth]{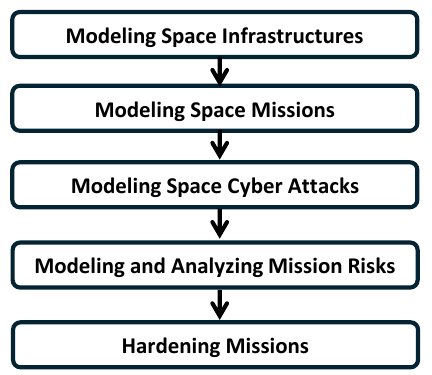}}
\caption{SCRM framework overview}
\label{fig:framework}
\end{figure}

\noindent{\bf SCRM Framework Overview}.
The targeted users of SCRM include:
(i) space infrastructure designers;
(ii) space infrastructure developers, who engineer countermeasures against space cyber threats;
(iii) space infrastructure defenders, who manage and operate cyber defenses;
(iv) space cyber risk management teams, who plan space missions and assure their success;
(v) researchers who want to improve our framework;
(vi) educators on
space cyber risk management;
and
(vii) trainers of space cyber risk management professionals.
Example use cases of SCRM include: (i) identifying and mitigating space cyber risks;
(ii) identifying and mitigating
space cyber risks to space infrastructure and missions, such as those imposed by identified vulnerabilities;
(iii) simulating or emulating space cyber risks in a testbed.
Figure \ref{fig:framework} highlights SCRM, which has five components and is elaborated below.

\subsection{Modeling Space Infrastructures}

\begin{figure}[!htbp]
\centering{\includegraphics[width=\columnwidth]{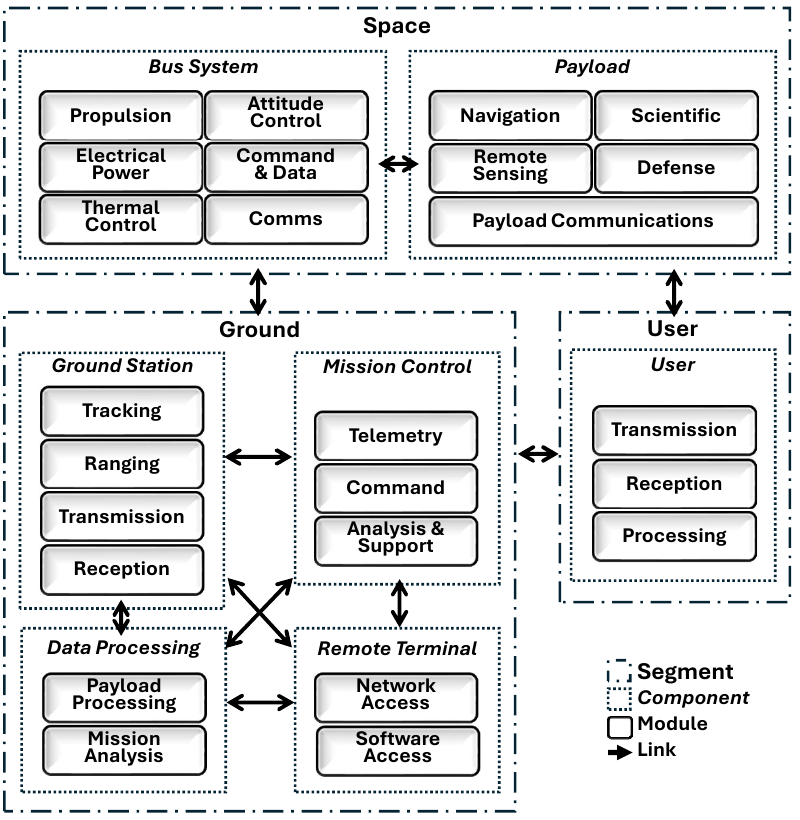}}
\caption{Space infrastructure system model}
\label{fig:space-model}
\end{figure}

Figure~\ref{fig:space-model} highlights our space infrastructure system model, which consists of four {\em segments} (i.e., ground, user, space, and link)
and is consistent with the widely used hierarchical segment-component-module levels of abstraction \cite{tedeschi2022satellite, guo2021survey, fritz2013satellite,XuSpaceSec2025} (we refer to these references for more detailed descriptions). Each segment has multiple {\em components}, with each accomplishing some sub-objectives.
Each component has some {\em modules}, with each accomplishing some of the component's sub-objectives. This leads to the following definition.

\begin{definition}
[space infrastructure representation]
\label{def:infrastructure representation}
A space infrastructure is modeled as a directed graph $G_{\general}=(V_{\general},E_{\general})$, where $V_{\general}$ is the set of nodes representing space infrastructure {\em modules}
and $E_{\general}$ is the set of arcs representing communication relationships such that $(u,v)\in E_{\general}$ means that $u\in V_{\general}$ can
{\em directly} communicate with $v_2\in V_{\general}$ (possibly via a communication path).
\end{definition}

Definition~\ref{def:infrastructure representation} focuses on the {\em module} level of abstraction, at which mission control and data flows are  described~\cite{XuIEEESP2025}, while aligning with NRS~\cite{sparta_notional,Ear2024NRS,XuIEEECSR2025-NRS}. However, the definition can be  adapted to describe space infrastructures at other levels of abstraction as needed. Note that $(u,v)\notin E_{\attackflow}$ implies that compromising node $u$ does not enable the attacker to wage a {\em direct} attack against node $v$ from node $u$, because $u$ cannot directly communicate with $v$ or the communication from $u$ to $v$ is filtered by a firewall-like mechanism. However, it is possible that compromise of $u$ directly leads to compromise of $w$ which further directly leads to compromise of $v$ because $(u,w)\in E_{\attackflow}$ and $(w,v)\in E_{\attackflow}$, manifesting the phenomenon of {\em cascading effects} that must be adequately treated.
$G_{\general}$ can be a multigraph when the communication $(u,v)\in E_{\general}$ can be implemented with different means or routing paths, such as Radio Frequency (RF) vs. Optical communications.

\subsection{Modeling Space Missions}
\label{sec:missions}

Given $G_{\general}=(V_{\general},E_{\general})$, we formalize the intuitive notions of mission {\em control flows} and {\em data flows} introduced in~\cite{XuIEEESP2025} as follows.

\begin{definition}[mission control flow]
\label{def:mcf}
A mission control flow
describes how
commands (or instructions) are issued between the nodes in $V_{\general}$. Formally, a mission control flow is represented as a {\em directed graph} $G_{\mcf}=\left(V_{\mcf},E_{\mcf}\right)$, where $V_{\mcf}\subseteq V_{\general}$, $E_{\mcf}\subseteq E_{\general}$, node $v\in V_{\mcf}$ represents a space infrastructure module, and arc $(u,v)\in E_{\mcf}$ indicates commands are issued from node $u$ to node $v$.
\end{definition}

Definition~\ref{def:mcf} indicates that a control flow is {\em not} necessarily a {\em line graph}, differing from the intuitive meaning of {\em flow} that one may think.
When the need arises, we use superscript $(i,j)$ to denote the $i$th control flow of the $j$th mission, leading to
$\left\{G^{(i,j)}_{\mcf}=\left(V^{(i,j)}_{\mcf},E^{(i,j)}_{\mcf}\right)\right\}$, where $1\leq i \leq m_j$ (i.e., mission $j$ has $m_j$ control flows) and $1\leq j \leq n$.

\begin{definition}[mission data flow]
\label{def:mdf}
A mission data flow describes how data items are transmitted between the nodes in $V_{\general}$. Formally, a mission data flow is a {\em directed graph} $G_{\mdf}=(V_{\mdf},E_{\mdf})$, where $V_{\mdf}\subseteq V_{\general}$, $E_{\mdf}\subseteq E_{\general}$,  $v\in V_{\mdf}$ represents a module, and arc $(u,v)\in E_{\mdf}$ indicates data items are sent from $u$ to $v$.
\end{definition}

Definition~\ref{def:mdf} also indicates that a data flow is {\em not} necessarily a {\em line graph} but an arbitrary directed graph, while it is possible $V_{\mcf}\cap V_{\mdf} \neq \emptyset$.
When the need arises, we use superscript $(i',j)$ to denote the $i'$th data flow of the $j$th mission,
$\left\{G^{(i',j)}_{\mdf}=\left(V^{(i',j)}_{\mdf},E^{(i',j)}_{\mdf}\right)\right\}$, where $1\leq i' \leq m'_j$ and $1\leq j \leq n$.

\subsection{Modeling Space Cyber Attacks}
\label{sec:modelingCyberAttacks}

\begin{definition}
[space cyber attack]
\label{definition:space-cyber-attack}A space cyber attack is any cyber attack that can cause damage to any of the four segments of a space infrastructure.
\end{definition}

Definition \ref{definition:space-cyber-attack} highlights that a space cyber attack does not necessarily directly target the space segment (e.g., satellite), but may target the ground or user segment while possibly pivoting to the space segment.
To model space cyber attacks, we adopt practitioners' way of thinking in modeling space cyber attack capabilities via {\em attack  techniques} as described in the SPARTA framework~\cite{sparta_homepage} and the MITRE ATT\&CK framework~\cite{ATTCK} (i.e., methods or approaches for achieving attacker sub-objectives), while noting that the former primarily describes cyber threats against the space and link segments and the latter primarily describes cyber threats against the ground and user segments (i.e., ATT\&CK is relevant according to Definition \ref{definition:space-cyber-attack}).
We choose to focus on these industrial standards because we hope this will ease the adoption of our framework in practice. Specifically, SPARTA specifies 9 space cyber {\em attack tactics} (i.e., an attacker's sub-objectives in different phases of attacks), with each supported by 5 to 18 attack techniques (some of which are further divided into 1 to 9 sub-techniques); ATT\&CK specifies 14 cyber attack tactics, with each supported by 8 to 44 techniques (some of which are further divided into 1 to 17
sub-techniques). This leads to:

\begin{definition}
[space cyber attack capabilities]
\label{def:attack capabilities}
We define the capabilities of space cyber attacker $A$ as the set of attack techniques
(as defined in ATT\&CK and SPARTA) that are possessed by $A$, denoted by $\AT_A$.
\end{definition}

\subsection{Modeling and Analyzing Mission Risks}

In order to adequately cope with space cyber risks, we must address space cyber attack cascading effects
from a mission perspective.

\subsubsection{Modeling Space Cyber Attack Cascading Effects}
\label{sec:cascading-effects}

We consider the following 3 kinds of {\em cascading effects} with respect to $e=(u,v)\in E_{\general}$.
\begin{itemize}
\item Type-1: Compromising node (i.e., module) $u$ can cause compromise of $v$. This can happen (e.g.) when $u$ is designed to send commands to $v$ (via a cryptographically authenticated channel) because a compromised $u$ can send malicious commands to $v$ for execution.
\item Type-2: Compromising arc or communication channel $(u,v)$ can cause the compromise of $v$. This can happen when the attacker injects malicious commands to $v$ for execution because, for instance, when the channel $(u,v)$ is not cryptographically authenticated.
\item Type-3: Compromising node $u$ can cause the compromise of $e=(u,v)$. This can happen (e.g.) when
compromising $u$ gives $A$ the privilege
to use $e$.
\end{itemize}
As elaborated later (Algorithm \ref{alg:space cyber risks analysis}), these cascading effects will be treated differently.
Note that there may or may not be other kinds of cascading effects, which are left to future studies.

\subsubsection{Modeling Space Cyber Risks} Risks are inherently associated with {\em uncertainty}, while noting that {\em certainty} can be seen as a special case of uncertainty (i.e., occurring with  likelihood 0 or 1). For instance, a space infrastructure may be attacked by attacker $A$ who {\em potentially} (or {\em certainly} as a special case) uses a specific attack technique, which can be described via likelihood $\alpha \in [0,1]$ (with $\alpha=1$ being the special case of certainty). We treat likelihood $\alpha$ as a broader concept than {\em probability} because $\alpha$ in the real world may be {\em subjectively} estimated by domain experts without explicitly considering the relevant probabilistic structure, as shown by SPARTA NRS~\cite{sparta_notional}. The following definition provides a framework, with a concrete example instantiation to illustrate ideas, for deriving the likelihood that a mission is disrupted from the likelihoods that the nodes and arcs associated with its controls flows and data flows are disrupted.

\begin{definition}
[mission disruption likelihood]
\label{definition:mission disruption likelihood}
Let $\L(\cdot)$ be a function returning the likelihood that a node, arc, control flow, data flow, or mission is disrupted by space cyber attacks.
We say mission $j$, which has $m_j$ control flows and $m'_j$ data flows specified via
$\left\{G^{(i,j)}_{\mcf}=\left(V^{(i,j)}_{\mcf},E^{(i,j)}_{\mcf}\right)\right\}$ and
$\left\{G^{(i',j)}_{\mdf}=\left(V^{(i',j)}_{\mdf},E^{(i',j)}_{\mdf}\right)\right\}$ where $1\leq i \leq m_j$, $1\leq i' \leq m'_j$, and $1\leq j \leq n$,
is disrupted with likelihood $\eta\in [0,1]$, denoted by $\L(j)=\eta$, if any of its control flow or data flow is disrupted at most with likelihood $\eta$, namely
\begin{equation}
\label{eq:L(j) aggregation function}
\begin{aligned}
\L(j)\gets\max\Bigl\{&\left\{\L\left(G^{(i,j)}_{\mcf}\right)\right\}_{1\leq i \leq  m_j}, \\
&\left\{\L\left(G^{(i',j)}_{\mdf}\right)\right\}_{1\leq i' \leq m'_j}\Bigr\}.
\end{aligned}
\end{equation}
\end{definition}

Definition~\ref{definition:mission disruption likelihood}  captures the intuition of ``weakest link'' in cybersecurity via the $\max$ function, which represents a specific method for deriving the likelihood that a mission is disrupted given the likelihoods that the nodes and arcs associated with its control flows and/or data flows are disrupted.
Nevertheless, other mathematical functions than the $\max$ function in Eq.\eqref{eq:L(j) aggregation function} are possible. That is, the definition can be adapted to accommodate an analyst's desired methods for deriving $\L\left(G_{\mcf}^{(i,j)}\right)$ from the $\L(v)$'s where $v\in V_{\mcf}^{(i,j)}$ and from the $\L(e)$'s where $e\in E_{\mcf}^{(i,j)}$, and for deriving $\L\left(G_{\mdf}^{(i',j)}\right)$ from the $\L(v)$'s where $v\in V_{\mdf}^{(i',j)}$ and from the $\L(e)$'s where $e\in E_{\mdf}^{(i',j)}$.

\subsubsection{Computing Space Cyber Risks to Missions}
We propose Algorithm~\ref{alg:space cyber risks analysis} for computing space cyber risks with respect to  missions. At a high level, its input includes a space infrastructure model, a set of attack techniques, a set of mission control flows, and a set of mission data flows. Its output is the mission disruption likelihood (Definition \ref{definition:mission disruption likelihood})
while considering the
{\em cascading} effects of space cyber attacks (Section \ref{sec:cascading-effects}).
The basic idea behind the algorithm can be understood as follows.
{\bf Lines 1-4}: Initialize {\em direct} and {\em cascading} compromise likelihoods.
{\bf Lines 5-11}: Compute the likelihood of attack techniques on nodes and arcs.
{\bf Lines 12-15}: Compute joint effect of direct attacks against nodes and arcs.
{\bf Lines 16-17}: Delete the nodes and arcs that cannot be attacked by the attack techniques {\em and} are not susceptible to cascading effects (if applicable); this is to make the algorithm widely applicable.
{\bf Lines 18-23}: Compute cascading effects on nodes and arcs.
{\bf Lines 24-26}: Compute mission disruption likelihood via a sub-routine (e.g., Algorithm \ref{alg:mission disruption likelihood}).

Now we look at the algorithm in greater detail.
Its input includes: a set of cyber attack techniques $\AT_A$ that may be possessed by an attacker $A$ with likelihood $\Lat_{\at}\in (0,1]$ for $\at\in \AT$, which may be derived from or provided by cyber threat intelligence (CTI); the likelihood that $\at$ can compromise node $v$, denoted by $\beta(v,\at)\in [0,1]$, or compromise arc $e$, denoted by $\beta(e,\at)\in [0,1]$, which may be obtained via CTI; space infrastructure $G_{\general}$; a set of $m_j$ mission control flows (indexed by $1\leq i\leq m_j$) and a set of $m_j'$ mission data flows (indexed by $1\leq i'\leq m_j'$) for mission $j$ where $1\leq j\leq n$; a $\flag\in \{0,1\}$, which allows the analyst to restrict the scope of
cascading effects (i.e., whether the nodes and arcs that cannot be compromised by attacker's attack techniques are susceptible to cascading effects or not).
It outputs $\L(j)$, the likelihood that mission $j$ is disrupted for $1\leq j \leq n$.

\begin{algorithm}[!htbp]
\small
\caption{Computing Space Cyber Risks}
\label{alg:space cyber risks analysis}
\KwInput{a set $\AT_A$ of attack techniques possessed by attacker $A$, associated with likelihood of possession $\Lat_{\at}\in (0,1]$ for $\at\in \AT_A$; the
likelihood $\at\in \AT_A$ can compromise node $v\in V_{\general}$, denoted by $\beta(v,\at)$, or arc $e\in E_{\general}$, denoted by $\beta(e,\at)$;
space infrastructure $G_{\general}=(V_{\general},E_{\general})$; a set of mission control flows $\left\{G^{(i,j)}_{\mcf}=\left(V^{(i,j)}_{\mcf},E^{(i,j)}_{\mcf}\right)\right\}$ and a set of mission data flows $\left\{G^{(i',j)}_{\mdf}=\left(V^{(i',j)}_{\mdf},E^{(i',j)}_{\mdf}\right)\right\}$ for $1\leq i \leq m_j$, $1\leq i' \leq m'_j$, and $1\leq j \leq n$; $\flag\in\{0,1\}$}

\KwOutput{mission disruption  likelihood $\L(j)$ for $1\leq j \leq n$}

$\L(v,\at)\gets 0$ for $v\in V_{\general}$ and  $\at \in \AT_A$

$\L(e,\at)\gets 0$ for $e\in E_{\general}$ and $\at \in \AT_A$

$\L(v)\gets 0$ for $v\in V_{\general}$

$\L(e)\gets 0$ for $e\in E_{\general}$

\For(~ {\tt // computing direct effects}){technique $\at\in \AT_A$}
    {
    \For{$v\in V_{\general}$}
    {
    \If{$\at$ compromises $v$ with likelihood $\beta(v,\at)>0$}
        {
    $\L(v,\at)\gets \beta(v,\at) \times \Lat_{\at}$
    }
    }
    \For{$e\in E_{\general}$}
    {
    \If{$\at$ compromises $e$ with likelihood $\beta(e,\at)>0$}
        {
    $\L(e,\at)\gets \beta(e,\at) \times \Lat_{\at}$
    }
    }
    }

    \For(~ {\tt // computing joint attack effects}){$v\in E_{\general}$}
    {
    $\L(v)\gets f(\{\L(v,\at)\}_{\at \in \AT_A})$ \tcp{using suitable $f$}
    }
    \For(~{\tt // using suitable $g$}){$e=(u,v)\in E_{\general}$}
    {
    $\L(e)\gets g(\L(u),\{\L(e,\at)\}_{\at \in \AT_A})$
    }

\If({\tt // Case 0 ($\flag=0$) vs. 1 ($\flag=1$)}){{$\flag=1$}}
{

delete $\{v:\L(v)=0 ~\text{and} \prod_{(u,v)\in E_{\general}}\L(u)=0\}$ and their adjacent arcs from $G_{\general}$

}

\Repeat (~{\tt // computing cascading effects}){$\{\L(v)\}_{v\in V_{\general}}$
converge (i.e., no longer updated)}
{
    \For(~ {\tt // using suitable $h$}){$v\in V_{\general}$}
    {
    $\L(v)\gets h(\{\L(u),\L(e=(u,v))\}_{e\in E_{\general}},\L(v))$

    \For(~{\tt // using suitable $h'$}){$e'=(v,v')\in E_{\general}$}
    {
    $\L(e')\gets h'(\L(v),\{\L(e',\at)\}_{\at \in \AT_A})$
    }
    }
}

\For{$j= 1 ~\text{to}~n$}
{
call a sub-routine (e.g., Algorithm~\ref{alg:mission disruption likelihood}) to compute mission disruption likelihood
$\L(j)$ based on $\{\L(v)\}_{v\in V_{\general}}$, $\{\L(e)\}_{e\in E_{\general}}$, $\left\{G^{(i,j)}_{\mcf}=\left(V^{(i,j)}_{\mcf},E^{(i,j)}_{\mcf}\right)\right\}_{1\leq i \leq m_j}$ and $\left\{G^{(i',j)}_{\mdf}=\left(V^{(i',j)}_{\mdf},E^{(i',j)}_{\mdf}\right)\right\}_{1\leq i' \leq m'_j}$
}

\Return{$\{\L(j)\}_{1\leq j \leq n}$, $\{\L(v)\}_{v\in V_{\general}}$, $\{\L(e)\}_{e\in E_{\general}}$}

\end{algorithm}

Lines 1-2 initialize $\L(v,\at)$ and $\L(e,\at)$, the likelihoods node $v\in V_{\general}$ and arc $e\in V_{\general}$ are compromised as a direct consequence of using attack technique $\at\in \AT_A$ to attack them.
Lines 3-4 initialize $\L(v)$ and $\L(e)$, the likelihoods $v\in V_{\general}$ and $e\in V_{\general}$ are compromised as a consequence of cascading effects and/or collective effect (i.e., multiple $\at$'s are applicable to them).

Lines 5-11 compute the {\em direct} effect of each attack technique $\at\in \AT_A$ on $v\in V_{\general}$ and $e\in V_{\general}$. The likelihood that attack technique $\at$ compromises $v$, denoted by $\L(v,\at)$, is the product of $\Lat_{\at}$, which is the likelihood that $\at$ is possessed by attacker $A$, and $\beta(v,\at)$, which is the likelihood that $\at$ compromises $v$.
The likelihood $\L(e,\at)$ that $e$ is compromised by attack technique $\at$ is similarly computed.

Lines 12-13 compute the {\em joint} effect of direct attacks that employ the applicable $\at \in \AT_A$ to attack node $v\in V_{\general}$ via a suitable
function $f$; note that here, and elsewhere in any algorithm, ``suitable" means a function must obey probability laws (i.e., they are probabilistic functions).
In the case that the successes of the $\at$'s are independent of each other, the likelihoods can be interpreted as independent probabilities and we can define function $f$ as:
\begin{equation}
\label{eq:function-f}
\L(v)\gets 1- \prod_{\at \in \AT_A} \left(1- \beta(v,\at) \times \Lat_{\at}\right).
\end{equation}
In the case of dependence, computing $\L(v)$ would require further information about the dependence structure (see, e.g.,~\cite{XuInternetMath2015Dependence} for a theoretical treatment of dependence in a related context); in the absence of such information, we suggest conducting experiments to empirically determine $\L(v)$
possibly by observing cascading effects of space cyber attacks in a space testbed, or asking domain experts to estimate them as in
\cite{sparta_notional,Ear2024NRS,XuIEEECSR2025-NRS}. Note that Eq.\eqref{eq:function-f} accommodates the worst-case scenario in the sense that we do not consider the possibility that an attack attempt is detected and then thwarted by the defender, which is left to future studies.

Lines 14-15 use an appropriate function $g$ to compute the {\em joint} effect of direct attacks waged from the applicable $\at \in \AT_A$ against arc $e\in V_{\general}$.
In the case that successes of these events are independent of each other, the likelihoods can be interpreted as independent probabilities and we define $g$ as:
\begin{equation}
\label{eq:function-g}
\L(e)\gets 1-
\prod_{\at \in \AT_A} \left(1- \beta(e,\at) \times \Lat_{\at}\right).
\end{equation}
In the case of dependence, $\L(e)$ can be treated in the same fashion as how $\L(v)$ would be treated. Similarly, Eq.\eqref{eq:function-g} also accommodates the worst-case scenarios that the attacks are not detected by defenses.

Line 16 uses a flag to indicate
whether to consider cascading effects
on the nodes and arcs that cannot be directly attacked by the techniques $\at \in \AT_A$. Specifically, $\flag=0$, also dubbed Case 0, means these
nodes and
arcs can be compromised as a consequence of
cascading effects (i.e., the compromise of an electrical power module increasing the likelihood that a communications module is compromised); otherwise (i.e. $\flag=1$, dubbed Case 1), these nodes and arcs can be deleted from $G_{\general}$ (line 17) as they need no further consideration.
 As we will show in our case study, contrasting Case 0 vs. Case 1 will lead to deep insights into the impact of cascading effects. Note that $\{v:\L(v)=0 ~\text{and} \prod_{(u,v)\in E_{\general}}\L(u)=0\}$ means the set of nodes that cannot be compromised by the techniques in $\AT_A$ and their parent nodes cannot be compromised by techniques in $\AT_A$.

Lines 18-23 compute the cascading effects of attacks via an suitable function $h$.
Similar to functions $f$ and $g$, the form of $h$ relies on whether the relevant events are independent of each other or not.
In the case of independence while assuming compromising $u$ can cause the compromise of $v$ (i.e., Type-1 cascading effects) and compromising $e=(u,v)$ can cause the compromise of $v$ (i.e., Type-2 cascading effects), the likelihoods can be interpreted as independent probabilities and we can define $h$ as:
\begin{equation}
\label{eq:function-h}
\L(v)\gets \L(v) + (1-\L(v))(1- (1-\L(u)) (1- \L(e))).
\end{equation}
Lines 21-22 update the likelihood that arc $e'=(v,v')$ is compromised as the likelihood $\L(v)$ is compromised has just been updated (i.e., Type-3 cascading effects). Under a similar independence assumption, we define $h'$ as:
\begin{equation}
\label{eq:function-h'}
\L(e')\gets \L(e') + (1-\L(e'))\L(v).
\end{equation}
Line 23 deserves special mention. While it is intuitive that the loop will eventually stop because {\em all vulnerable nodes are eventually compromised with a respective likelihood that depends on the structure of $G_{\general}$ in the absence of defense}, e.g., a node $v$ with in-degree 0 is compromised with likelihood $\L(v)$ of direct attacks without being affected by any cascading effect.
A rigorous proof of this property is left to future studies, but could be obtained by leveraging the theoretical results reported in
~\cite{XuIEEETNSE2018,XuIEEEACMToN2019}, which rigorously prove that a large family of cybersecurity dynamics is globally convergent, prompting Line 23
always converges.

Lines 24-26 call a sub-routine to compute the likelihood that mission $j$ is disrupted, where $1\leq j \leq n$. There can be multiple methods for this purpose, such as the one described in Algorithm~\ref{alg:mission disruption likelihood}, which essentially uses the $\max$ function to aggregate attack effects (corresponding to
the cybersecurity intuition of ``weakest link'').

The computational complexity of Algorithm~\ref{alg:space cyber risks analysis} is estimated as follows. The first loop (Lines 5-11) incurs a complexity of $O(|\AT_A| \times |V_{\general}|^2)$, assuming there are at most a constant number of multi-arcs between any pair of nodes (which would be true in practice). This is also the upper bound of the number of calls to function $f$ (Lines 12-13), of the number of calls to function $g$  (Lines 14-15), and of Lines 16-17 which only incurs $O(|V_{\general}|^2)$.
However, it is difficult to estimate how many iterations it takes the cascading effects to converge (Lines 18-23) because the convergence result,  even if theoretically proven (as shown in~\cite{XuIEEETNSE2018,XuIEEEACMToN2019}), is asymptotic. In our experiments, they converge quickly. The last loop (Lines 24-26) incurs a complexity of $n$ times the complexity of Algorithm~\ref{alg:mission disruption likelihood}.

\subsubsection{Computing Mission Disruption Likelihood}

\begin{algorithm}[!htbp]
\small
\caption{Sub-routine for computing $\L(j)$, the likelihood that mission $j$ is disrupted by attacker $A$}\label{alg:mission disruption likelihood}
\KwInput{$\{\L(v)\}_{v\in V_{\general}}$,  $\{\L(e)\}_{e\in E_{\general}}$,
$\left\{G^{(i,j)}_{\mcf}=\left(V^{(i,j)}_{\mcf},E^{(i,j)}_{\mcf}\right)\right\}_{1\leq i \leq m_j}$ and $\left\{G^{(i',j)}_{\mdf}=\left(V^{(i',j)}_{\mdf},E^{(i',j)}_{\mdf}\right)\right\}_{1\leq i' \leq m'_j}$
}

\KwOutput{$\L(j)$, the likelihood mission $j$ is disrupted}

\For{$i=1~\text{to}~m_j$}
{
compute the likelihood that the $i$th control flow of mission $j$'s is disrupted as
$\L\left(G_{\mcf}^{(i,j)}\right)\gets{\max}\left\{{\max}_{v\in V^{(i,j)}_{\mcf}} \L(v),\max_{e\in E^{(i,j)}_{\mcf}} \L(e)\right\}$
}

\For{$i'=1~\text{to}~m'_j$}
{
compute the likelihood that the $i'$th data flow of mission $j$'s is disrupted as $\L\left(G_{\mdf}^{(i',j)}\right)\gets{\max}\left\{{\max}_{v\in V^{(i',j)}_{\mdf}} \L(v),\max_{e\in E^{(i',j)}_{\mdf}} \L(e)\right\}$
}

compute the likelihood mission $j$ is disrupted via Eq.\eqref{eq:L(j) aggregation function} or $\L(j)\gets \max\left\{\left\{\L\left(G_{\mcf}^{(i,j)}\right)\right\}_{1\leq i \leq m_j}, \left\{\L\left(G_{\mdf}^{(i',j)}\right)\right\}_{1\leq i' \leq m'_j}\right\}$

\Return{$\L(j)$}

\end{algorithm}

Algorithm~\ref{alg:mission disruption likelihood} is a specific method for computing mission disruption likelihood $\L(j)$, which uses the $\max$ function to accommodate the the cybersecurity notion of ``weakest link.''
This $\max$
function considers that each mission control flow and mission data flow is essential to the success of the mission. As such, the disruption likelihood of a single mission control flow or mission data flow equates to the disruption likelihood of the mission.
For a given mission control flow, its disruption likelihood is defined as the maximum likelihood among the $\L(v)$'s for $v\in V_{\mcf}$ and the $\L(e)$'s for $e\in E_{\mcf}$ (Lines 1-2). For a given mission data flow, its disruption likelihood is similarly computed (Lines 3-4). The mission's disruption likelihood is the maximum among the control flow disruption likelihoods and the data flow disruption likelihoods (Line 5).

The complexity of Algorithm~\ref{alg:mission disruption likelihood} is $O\left((m+m')|V_{\general}|^2\right)$, where $m=\max\{m_1,\ldots,m_n\}$ and $m'=\max\{m'_1,\ldots,m'_n\}$, while noting $G^{(i,j)}_{\mcf}\subseteq G_{\general}$ and $G^{(i,j)}_{\mdf}\subseteq G_{\general}$ and assuming there is a bounded
number of arcs between any pair of nodes.

\subsection{Hardening Space Missions}

Let $\tau\in [0,1]$ denote the tolerable degree of mission disruption. If the output of Algorithm~\ref{alg:space cyber risks analysis} is greater than $\tau$, namely $\L(j)>\tau$ for some $1\leq j \leq n$, then {\em security controls} (i.e., countermeasures) need to be identified and employed to reduce the mission disruption likelihood to a tolerable level (i.e., below $\tau$). This triggers the need for a {\em mission hardening} algorithm, such as Algorithm~\ref{alg:mission hardening},
for identifying security controls, denoted by $\dt$'s, that can be employed to harden missions. For this purpose, we leverage the security controls specified in NIST SP 800-53~\cite{NIST800-53r5} as they are widely used.

\begin{algorithm}[!htbp]
\small
\caption{Hardening Missions $j$ ($1\leq j \leq n$)}
\label{alg:mission hardening}
\KwInput{tolerable mission disruption likelihood threshold $\tau\in [0,1]$; a set $\AT_A$ of attack techniques possessed by attacker $A$, associated with likelihood of possession $\Lat_{\at}\in (0,1]$ for $\at\in \AT_A$; the
likelihood that attack technique $\at\in \AT_A$ can compromise node $v\in V_{\general}$, denoted by $\beta(v,\at)$, or arc $e\in E_{\general}$, denoted by $\beta(e,\at)$;
space infrastructure $G_{\general}=(V_{\general},E_{\general})$;
a set of mission control flows $\left\{G^{(i,j)}_{\mcf}=\left(V^{(i,j)}_{\mcf},E^{(i,j)}_{\mcf}\right)\right\}$ and a set of mission data flows $\left\{G^{(i',j)}_{\mdf}=\left(V^{(i',j)}_{\mdf},E^{(i',j)}_{\mdf}\right)\right\}$ for $1\leq i \leq m_j$, $1\leq i' \leq m'_j$, and $1\leq j \leq n$;
$\{\L(j)\}_{1\leq j \leq n}$, $\{\L(v)\}_{v\in V_{\general}}$, $\{\L(e)\}_{e\in E_{\general}}$ (output of Algorithm~\ref{alg:space cyber risks analysis})
}

\KwOutput{security controls for mitigating risks to missions}

$\AT^*_A\gets \AT_A$

$J\gets \{j: (1\leq j \leq n)~\text{and}~(\L(j)>\tau)\}$

\If{$J= \emptyset$}
{
\Return{``Hardening not necessary''}
}
\Else{

\For(~{\tt // using updated $G_{\general}$ and $\AT_A$}){$v\in V_{\general}$}
{
If{~$\exists \at\in \AT_A$ such that $\beta(v,\at)\times \Lat(\at)>\tau$}
{
$\AT_A\gets \AT_A - \{\at\}$;  \hspace{1em} $ V_{\general}\gets V_{\general}-\{v\}$
}
}
\For(~{\tt // using updated $G_{\general}$ and $\AT_A$}){$e\in E_{\general}$}
{
If{~$\exists \at\in \AT_A$ such that $\beta(e,\at)\times \Lat(\at)>\tau$}
{
$\AT_A\gets \AT_A - \{\at\}$; \hspace{1em} $E_{\general}\gets E_{\general}-\{e\}$
}
}

call Algorithm~\ref{alg:space cyber risks analysis} with $\flag=0$ and the updated $G_{\general}$ and $\AT_A$  to re-compute $\L(j)$ with cascading effects

$J\gets \{j: (1\leq j \leq n)~\text{and}~(\L(j)>\tau)\}$

\While{$J\neq \emptyset$}
{
\For(~{\tt // using updated $G_{\general}$ and $\AT_A$}){$e\in V_{\general}$ such that $\L(e=(u,v))>\tau$}
{
$\AT_A\gets \AT_A-\{\at: \beta(u,\at)>0\}$

delete $u$ and its adjacent arcs from $G_{\general}$;

call Algorithm~\ref{alg:space cyber risks analysis} with $\flag=0$ and updated $G_{\general}$ and $\AT_A$ to re-compute $\L(j)$ to accommodate cascading effects;

$J\gets \{j: (1\leq j \leq n)~\text{and}~(\L(j)>\tau)\}$
}
}

\For{$\at\in \AT^*_A - \AT_A(v)$}
    {
    use sub-routine $\Delta$ to identify security control to thwart $\at$, denoted by $\dt(\at)$ at the relevant nodes and arcs
    }

\Return{security controls $\{\dt(\at)\}_{\at\in \AT^*_A-\AT_A}$}
}
\end{algorithm}

Algorithm~\ref{alg:mission hardening} determines whether mission hardening is required, and if so, how. It can be understood as follows. Line 1 keeps the input $\AT_A$ in $\AT^*_A$ because $\AT_A$ will be dynamically updated. Hardening is required  when there is some mission $j$ ($1\leq j \leq n$) that is disrupted with a likelihood $\L(j)>\tau$, as shown in Line 3. In the case of hardening (Lines 6-20), the algorithm proceeds as follows. The loop in Lines 6-7 identify the $\at$'s that must be mitigated because they would immediately disrupt a mission via some {\em nodes} even without considering the joint or cascading attack effects. The loop in Lines 8-9 identify the $\at$'s that must be mitigated because they would immediately disrupt a mission via some {\em arcs} even without considering the joint or cascading attack effects.
After mitigating these $\at$'s, Line 10 calls Algorithm~\ref{alg:space cyber risks analysis} to determine whether some mission is still disrupted, where $\flag=0$ because Algorithm~\ref{alg:mission hardening} will solely determine which $\at$'s to mitigate (while accommodating the worst-case scenarios that the nodes and arcs that cannot be directly attacked by $\at\in \AT_A$ are affected by
attack cascading effects). If so, the loop in Lines 12-17 further mitigates the remaining $\at$'s that contributes to the cascading effects.
The loop in Lines 18-19 identifies the NIST security controls that can mitigate the $\at\in \AT^*_A-\AT_A$ as identified by the algorithm.

The computational complexity of Algorithm~\ref{alg:mission hardening} is analyzed as follows. The loop in Lines 6-7 incurs an $O(|\AT_A|\times |V_{\general}|)$ complexity. The loop in Lines 8-9 incurs an $O(|\AT_A|\times |V_{\general}|^2)$ complexity, while assuming the number of arcs between any pair of nodes is bounded by a constant (which is often true in practice). The loop in Lines 12 repeats at most $|\AT_A|$ times, with the inner loop in Lines 13-17 for at most $O(|V_{\general}|^2)$ calls to Algorithm~\ref{alg:space cyber risks analysis}, leading to at most $O(|\AT_A| \times |V_{\general}|^2)$ calls  to Algorithm~\ref{alg:space cyber risks analysis}. The loop in Lines 18-19 incurs $O(|\AT_A|)$ calls to the sub-routine $\Delta$, which could be implemented with a simple table-querying algorithm (reflecting our manual process in our case study). Thus, the total complexity is $O(|\AT_A| \times |V_{\general}|^2)$ calls  to Algorithm~\ref{alg:space cyber risks analysis}.

\section{Case Study}\label{sec:caseStudy}

To demonstrate the {\em use} (owing to its complexity) and usefulness of SCRM, we conduct a case study based on our SATCOM (satellite communications) testbed.
We re-implement 3 real-world space cyber attacks against SATCOM (learned from authors of
\cite{ear2023characterizing}) and mitigate them via
Algorithm~\ref{alg:mission hardening}.

\subsection{Modeling Space Infrastructure (Testbed)}\label{subsec:TestbedInfrastructure}

Figure~\ref{fig:testbed} highlights our testbed, which includes 4 segments (i.e., space, link, ground, and user).

\begin{figure}[!htbp]
\setcounter{figure}{2}
\centering{\includegraphics[width=.6\columnwidth]{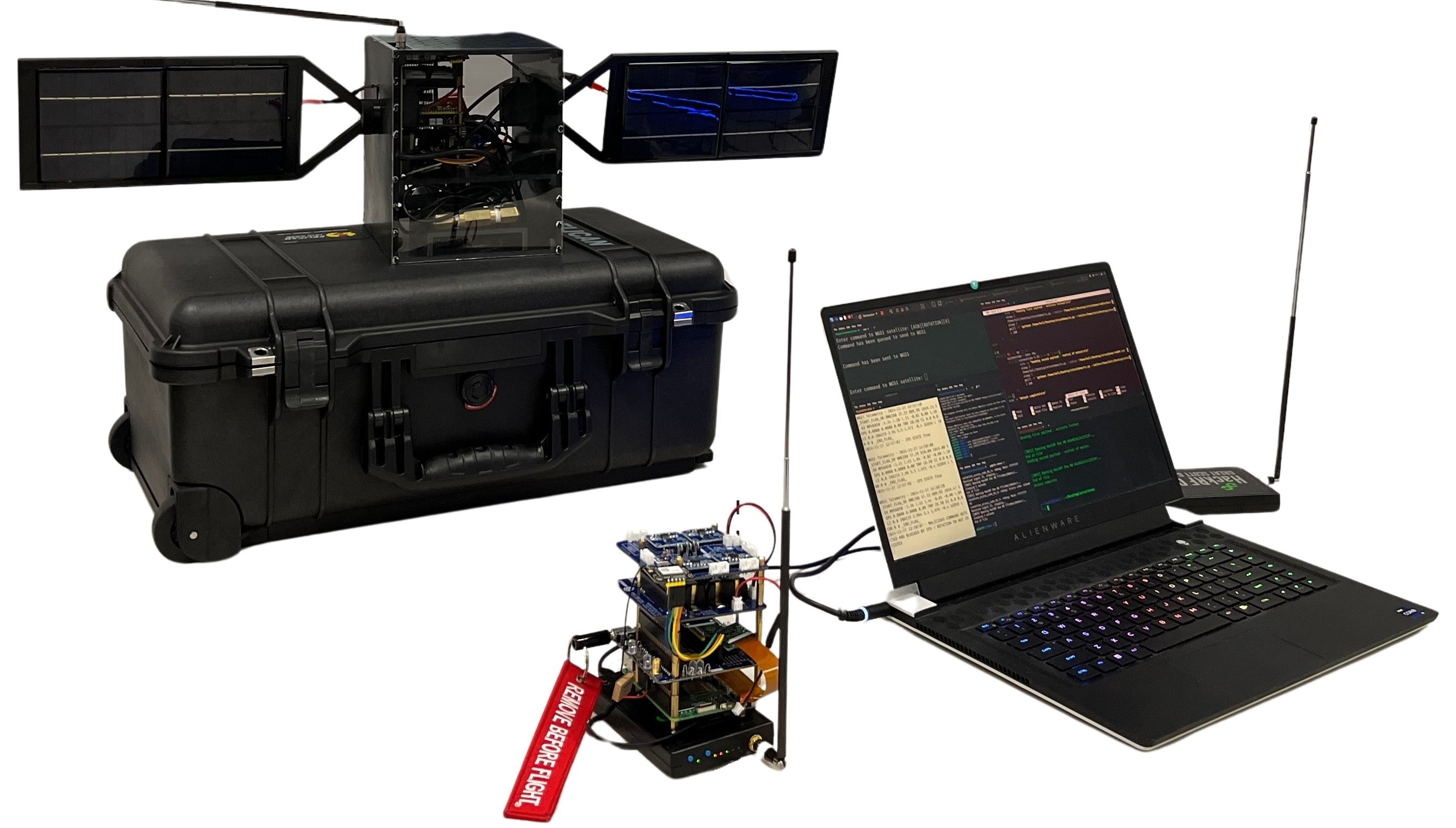}}
\caption{Our SATCOM space infrastructure testbed.}
\label{fig:testbed}
\end{figure}

\begin{figure}[!htbp]
\centering{\includegraphics[width=\columnwidth]{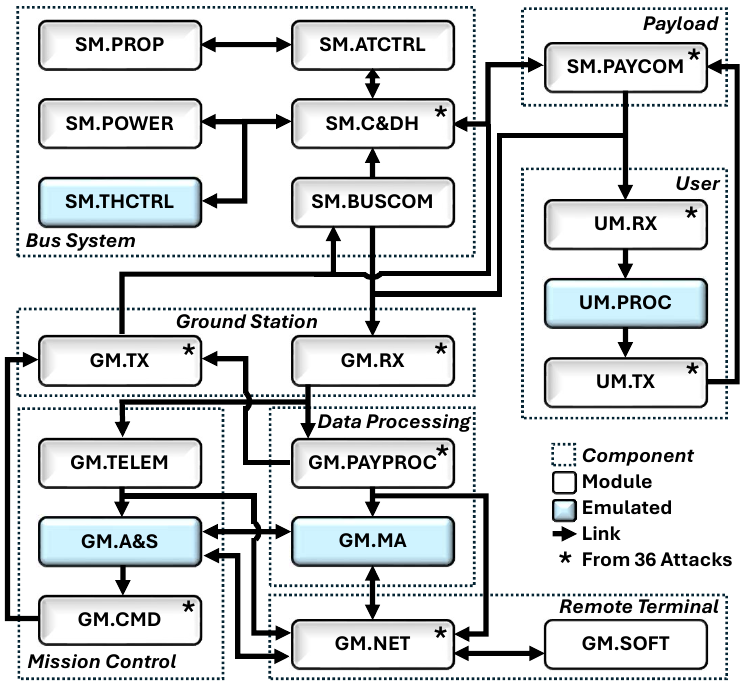}}
\caption{$G_{\general}$ of our testbed, with  $|V_{\general}|=19$ modules.} \label{fig:infrastructureGraph}
\end{figure}

Figure~\ref{fig:infrastructureGraph} describes the $G_{\general}$ corresponding to our testbed at the module-level, with $|V_{\general}|=19$ modules in total. The modules are named as follows: SM.NAME denotes a module in the Space segment, GM.NAME denotes a module in the Ground segment, and UM.NAME denotes a module in the User segment.
The space segment consists of 2 components: bus system (6 modules) and payload (one module); the payload or application mission is to enable two users to communicate with each other via the satellite.
The ground segment consists of 4 components: ground station (2 modules), mission control (3 modules), data processing (2 modules), and remote terminal (2 modules). The user segment has 1 component: user (3 modules). The link segment is indicated by bold arrows. The modules indicated with an `*' are involved in the aforementioned 36 real-world space cyber attacks;
these 9 modules are also involved in the mission that is used in our case study.
Among the 19 modules, 15 are fully
implemented and the other 4 are emulated (highlighted in light blue color) because these 4 involve
proprietary hardware and software that we do not have access to at the time of writing.

The 15 fully implemented modules are briefly described as follows.
In the ground segment, we employ Ball Aerospace's Comprehensive Open-architecture Solution for Mission Operations Systems (COSMOS) ground software and a Debian OS on an Intel i7 processor to implement GM.TELEM (Telemetry), GM.PAYPROC (Payload Processing), and GM.CMD (Command).
We employ Netgate's pfSense appliance and an OpenVPN service to implement GM.NET (Network Access) and GM.SOFT (Software Access).
In the Space segment, we employ NASA's core Flight System (cFS) flight software and a Debian OS on an ARM Cortex-A76 processor to implement SM.C\&DH (Command \& Data Handling), while implementing SM.ATCTRL (Attitude Control) via custom firmware and an ARM Cortex-M0+ processor on a RP2040 microcontroller.
We implement SM.PROP (Propulsion) with an electric relay, solenoid valve, $CO_2$ canister, and exhaust ports.
We implement SM.POWER (Electrical Power) with electric relays, servomotors, solar panel arrays, Texas Instruments' INA219 current and voltage sensors, and batteries.
In the Ground, Space, and User segments, we use GNU Radio libraries and Great Scott Gadgets' HackRFs to implement GM.TX (Ground Transmission), GM.RX (Ground Reception), UM.TX (User Transmission), UM.RX (User Reception), SM.PAYCOM (Payload Communications), and SM.BUSCOM (Bus Communications).

The other 4 modules are emulated as follows.
In the Ground segment, we emulate GM.A\&S (Analysis \& Support) and GM.MA (Mission Analysis) with Bash and Python scripts for input/output interactions.
In the Space segment, we emulate SM.THCTRL (Thermal Control) by producing ambient temperature, barometric pressure, and relative humidity measurements with Bosch Sensortec's BME280 sensor.
In the User segment, we emulate UM.PROC (User Processing) with Bash and python scripts.

\subsection{Modeling Space Missions (Testbed)}

The testbed currently supports the SATCOM mission, meaning $n=1$, with
$m_1=3$ control flows and $m'_1=2$ data flows.

\subsubsection{Bus Management Control Flow $G^{(1,1)}_{\mcf}$}

\begin{figure}[!htbp]
\centering{\includegraphics[width=\columnwidth]{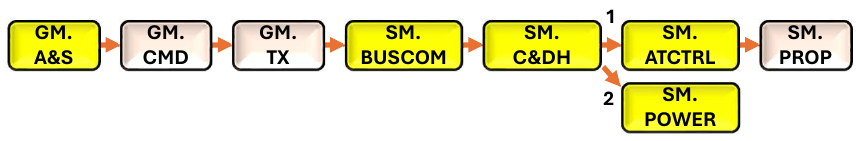}}
\caption{Bus Management Mission Control Flow $G^{(1,1)}_{\mcf}$.}
\label{fig:busMgmtGraph}
\end{figure}

Figure~\ref{fig:busMgmtGraph} depicts this  flow, which allows for commanding the modules of the bus component of the satellite.
More specifically, this control flow is for purposes such as:  (a) maneuvering the satellite to modify the pitch, roll, or yaw of the satellite to ensure optimal communications antenna position, or to adjust the satellite's orbit to meet the SATCOM's mission requirements; and
(b)
managing the electrical power distribution of the satellite,
such as powering off certain components
to preserve power or controlling the direction of the solar panel arrays.

For purpose (a), the control flow begins at module GM.A\&S (Analysis \& Support)  of the Mission Control component
and ends at module SM.PROP (Propulsion) of the Bus component.
It progresses as follows:
(i) GM.A\&S
determines whether a change in the satellite's pitch, roll, yaw, or orbit is required, and if so, sends the required state change (of Propulsion module SM.PROP) to module GM.CMD (Command);
(ii) GM.CMD
interprets and encodes the change requirement into the Consultative Committee for Space Data Systems (CCSDS) protocol, and sends the command data packet to the Transmission module GM.TX;
(iii) GM.TX
modulates the command data packet for Radio Frequency (RF) transfer
to the Communications module SM.BUSCOM of the Bus component;
(iv) SM.BUSCOM
demodulates the RF signal to retrieve the command data packet and forwards it to the Command \& Data Handling module SM.C\&DH;
(v) SM.C\&DH
parses, validates, and translates the command data packet to command directives for module SM.ATCTRL (Attitude Control);
(vi) SM.ATCTRL
parses and executes the command directives to send the corresponding electric signals to module SM.PROP (Propulsion); and
(vii) SM.PROP
receives the electric signals and actuates the solenoid vales to control propellant discharge to achieve the desired satellite maneuver.

For purpose (b),
GM.A\&S determines to change the power distribution or solar panel direction.
Eventually, module SM.C\&DH (Command \& Data Handling) pushes the power re-configuration command onto the flight software bus to make module SM.POWER (Power) to retrieve the command and execute it to manage the satellite's electrical power.

\subsubsection{Remote Management Control Flow $G^{(2,1)}_{\mcf}$}

\begin{figure}[!htbp]
\centering{\includegraphics[width=.41\columnwidth]{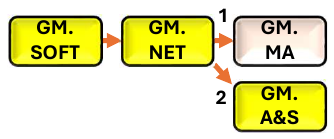}}
\caption{Remote Management Mission Control Flow $G^{(2,1)}_{\mcf}$.}
\label{fig:remoteMgmtGraph}
\end{figure}

Figure~\ref{fig:remoteMgmtGraph} depicts this control flow, which allows operators to remotely control the ground segment when away from the main Mission Control and Data Processing facility. The purposes include:
(a) for operators to establish, modify, or delete mission analysis parameters to ensure SATCOM mission success, such as
collecting channel saturation levels; and
(b) for operators to establish, modify, or delete infrastructure analysis parameters to ensure optimal satellite health, such as setting threshold levels for triggering sensor alerts.

For purpose (a), the control flow begins at the Software Access module GM.SOFT and ends at the Mission Analysis module GM.MA. It progresses as follows:
(i) the GM.SOFT
module enables an operator to compose and transmit configuration directives through a client interface that is connected to the GM.NET module via a VPN service that traverses commercial networks;
(ii) the Network Access module GM.NET
transmits the remote operator's configuration directives to the GM.MA module on a local LAN; and
(iii) the GM.MA
module receives and executes the new configuration directives to modify its analysis parameters for the SATCOM mission.

For purpose (b),
the control flow also begins at GM.SOFT but ends at module GM.A\&S (Analysis \& Support).
Eventually,
module GM.A\&S  receives the configuration directives from module GM.NET (Network Access) and executes it to modify its parameters for the SATCOM infrastructure.

\subsubsection{Subscription Management Control Flow $G^{(3,1)}_{\mcf}$}

\begin{figure}[!htbp]
\centering{\includegraphics[width=\columnwidth]{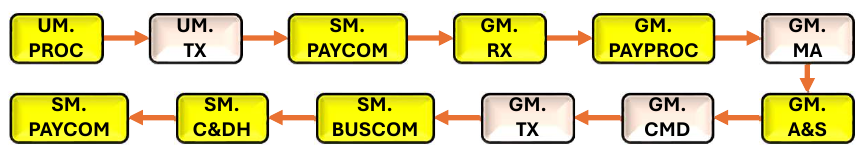}}
\caption{Subscription Mgmt Mission Control Flow $G^{(3,1)}_{\mcf}$.}
\label{fig:subscriptionMgmtGraph}
\end{figure}

Figure~\ref{fig:subscriptionMgmtGraph} depicts
this control flow for managing the channel subscription configuration of the satellite. It
allows SATCOM users to establish, modify, or delete their channel subscription, such as increasing the bandwidth of current subscription.
It contains 12 nodes, beginning at module UM.PROC (User Processing) and ending with  module SM.PAYCOM (Payload Communications).
It progresses as follows:
(i) UM.PROC receives a users' subscription change request and forwards it to module UM.TX (Transmission);
(ii) UM.TX
sends the change request to SM.PAYCOM via RF;
(iii) SM.PAYCOM acts as a repeater, forwarding the request to module GM.RX (Ground Reception) via RF;
(iv) GM.RX transmits the request to module GM.PAYPROC (Ground Payload Processing);
(v) GM.PAYPROC attaches the context of the users' current subscription to the change request
and forwards it to module GM.MA (Mission Analysis);
(vi) GM.MA
assesses if the change is allowed within the mission context, and if so, sends the request to module GM.A\&S (Analysis \& Support);
(vii) GM.A\&M assesses if the change is allowed,
and if so, sends the request to  module GM.CMD (Command);
(viii) GM.CMD
converts the change request into standard CCSDS commands and forwards the commands to module GM.TX (Transmission);
(ix) GM.TX
modulates the command data packet and sends the signal to module SM.BUSCOM (Communications) via RF;
(x) SM.BUSCOM
demodulates the RF signal and forwards the command data packet to module SM.C\&DH (Command \& Data Handling);
(xi) SM.C\&DH parses and validates the command data packet and sends the command to module SM.PAYCOM (Payload Communications);
(xii) SM.PAYCOM executes the commands to modify its channel subscription hardware and software configuration to meet the users' requirements.

\subsubsection{Telemetry Reporting and Analysis Data Flow $G^{(1,1)}_{\mdf}$}

\begin{figure}[!htbp]
\centering{\includegraphics[width=\columnwidth]{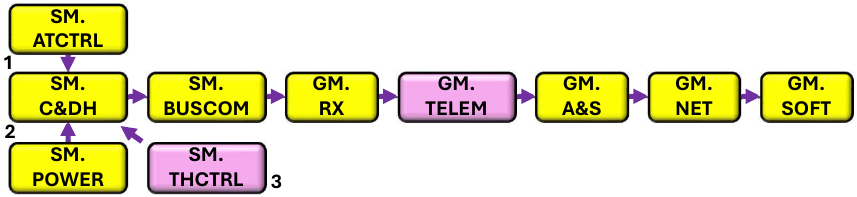}}
\caption{Telemetry Mission Data Flow $G^{(1,1)}_{\mdf}$.}
\label{fig:telemGraph}
\end{figure}

Figure~\ref{fig:telemGraph} depicts this flow,
which allows for transmitting satellite health and status data
to operators in the Ground segment.
More specifically, it serves 3 purposes:
(a)
transmitting sensor data about the pitch, roll, yaw of the satellite, as well as its propellant status (e.g., positions of relays, solenoid valves, and rates of propellant releases) to satellite operators at the Analysis \& Support module GM.A\&S or the Software Access module GM.SOFT;
(b) transmitting sensor data about the satellite's electric power (e.g., currents and voltages) and the servomotor angle position for the solar panel arrays; and
(c) transmitting sensor data about the satellite's thermal environment,
such as the temperature of the flight computer.

For purpose (a), the data flow
begins at the Attitude Control module SM.ATCTRL and ends at the Analysis \& Support module GM.A\&S.
It progresses as follows:
(i) SM.ATCTRL
pushes its sensor data onto the flight software bus;
(ii) the Command \& Data Handling module SM.C\&DH
periodically retrieves the sensor data, packs them as an archive,
and sends the archive to the Communication module SM.BUSCOM;
(iii) SM.BUSCOM
modulates the archive and sends the data signal via RF to the Ground Reception module GM.RX;
(iv) GM.RX
receives and demodulates the RF signal into the data archive and forwards it to the Telemetry module GM.TELEM;
(v) GM.TELEM
unpacks the data archive and populates the sensor data on its dashboards to monitor real-time satellite health and then forwards the sensor data to the Analysis \& Support module GM.A\&S;
(vi) GM.A\&S
analyzes the sensor data to determine trends in satellite health and forwards the data to the Network Access module GM.NET if there is a remote operator;
(vii) GM.NET transmits the sensor data through its VPN service to the Software Access  module GM.SOFT; and
(viii) GM.SOFT displays the sensor data in a client interface for the remote operator to examine.

For purpose (b),
it begins at module SM.POWER (Power) and ends at module GM.SOFT (Software Access).
Specifically, SM.POWER pushes its sensor data onto the flight software bus where module SM.C\&DH (Command \& Data Handling) periodically retrieves the sensor data.
The process for the remaining 7 nodes is the same as for purpose (a).

For purpose (c), it
begins at
SM.THCTRL and ends at module GM.SOFT (Software Access).
SM.THCTRL pushes its sensor data onto the flight software bus. The process for the remaining 7 nodes is the same as for purpose (a).

\subsubsection{Channel Operations Data Flow $G^{(2,1)}_{\mdf}$}

\begin{figure}[!htbp]
\centering{\includegraphics[width=.8\columnwidth]{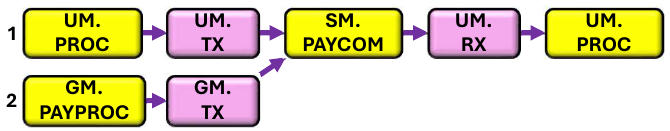}}
\caption{Channel Operations Mission Data Flow $G^{(2,1)}_{\mdf}$.}
\label{fig:channelGraph}
\end{figure}

Figure~\ref{fig:channelGraph} depicts this flow,
which is for transmitting data from its source to end-users. Specifically, this control flow serves 2 purposes:
(a)
transmitting communications data from one user to another user via a satellite to accomplish beyond-line-of-sight voice and data communications; and
(b)
transmitting communication data from the Ground segment to end-users (e.g., satellite television broadcasting).

For purpose (a), the control flow
begins at one user's module UM.PROC (Processing) and ends at another user's UM.PROC.
It progresses as follows:
(i) the sending user's UM.PROC module receives voice or data and packs the byte stream in the CCSDS packet structure;
(ii) the sending user's Transmission module UM.TX modulates the packets and transmits the signal via RF to the Bus Communications module SM.BUSCOM;
(iii) SM.BUSCOM
receives the data signal and acts as a repeater by re-transmitting the signal to the receiving user's Reception module UM.RX;
(iv) the receiving user's UM.RX
demodulates the data signal and sends the resulting data to the receiving user's Processing module UM.PROC; and
(v) the receiving user's UM.PROC presents the
data to the end-user via a client terminal interface.

For purpose (b),
the control flow begins at the Ground Payload Processing module GM.PAYPROC and ends at the User Processing module UM.PROC.
Specifically, GM.PAYPROC packages the communications data according to the CCSDS standard and the
Ground Transmission module GM.TX modulates the packets into a data signal.
From there, the remaining
process is the same as for (a).

\subsection{Modeling Space Cyber Attacks}\label{subsec:attacks}

We repeat and model 3 real-world space cyber attacks (learned from authors of
\cite{ear2023characterizing}). Attack 1 is the 2008 space cyber attack~\cite{fritz2013satellite} that used a malware to exploit an unpatched module in the Ground segment at the Johnson Space Center; the malware further leverages an uplink to gain control the International Space Station (ISS) in the Space segment. Attack 2 is the 1998 incident, originally reported in~\cite{fritz2013satellite}, that caused the on-board processor of the PANAMSAT Galaxy 4 SATCOM satellite to malfunction, leading to
a four-day disruption of 90\% of US pagers.
Attack 3 is
the 2007
space cyber attack, originally reported in~\cite{fritz2013satellite}, that
hijacked a Czech
satellite TV channel to transmit its own audio and video.

\subsubsection{Attacker Capabilities (Attacks 1-3)}
\label{sec:case study attacker capabilities}

\begin{figure}[!htbp]
\centering{\includegraphics[width=\columnwidth]{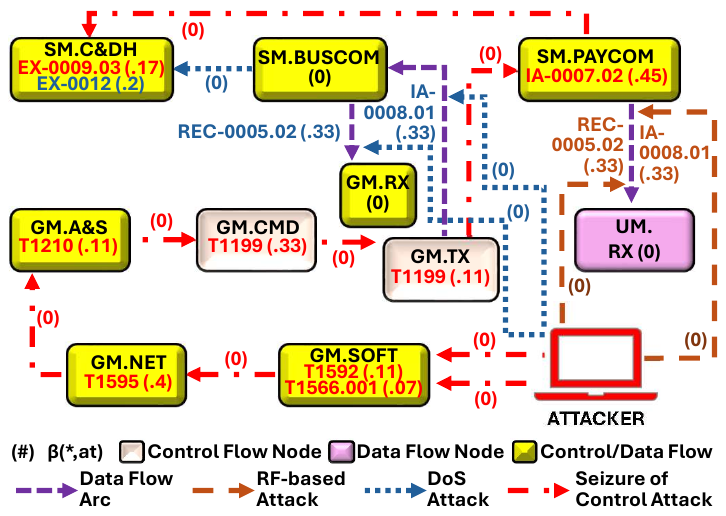}}
\caption{The 10 techniques
employed by  Attacks 1-3.}
\label{fig:attackFlowGraph}
\end{figure}

As highlighted in Figure~\ref{fig:attackFlowGraph}, Attacks 1-3 use 10 attack techniques ($|\AT_A|=10$) in total.
The 10 attack techniques are $\AT_A=
\{$T1210, T1199, T1595, T1592, T1566.001, EX-0012, EX-0009.03, IA-0007.02, IA-0008.01, REC-0005.02$\}$, with likelihood of possession by $A$,
namely $\Lat_{\at}=\{$0.23, 0.38, 0.38, 0.15, 0.25, 0.24, 0.23, 0.27, 0.23, 0.23$\}$.
The 10 techniques are:
(i) {\em Downlink Intercept} (SPARTA REC-0005.02), whereby an attacker eavesdrops on data signal broadcasts from a satellite;
(ii) {\em Rogue Ground Station} (SPARTA IA-0008.01), whereby an attacker injects their own ground station into a space infrastructure;
(iii) {\em Modify On-Board Values} (SPARTA EX-0012), whereby an attacker modifies either stored data or the configuration settings of modules in the space segment to alter or disrupt satellite behavior;
(iv) {\em Gather Victim Host Information} (ATT\&CK
T1592), whereby an attacker identifies important information about a victim host, such as operating system version;
(v) {\em Spearphishing Attachment} (ATT\&CK T1566.001), whereby an attacker convinces a victim to click on a malicious file;
(vi) {\em Active Scanning} (ATT\&CK T1595), whereby an attacker identifies important information about a victim network, such as open ports of connected hosts;
(vii) {\em Exploitation of Remote Services} (ATT\&CK T1210), whereby an attacker gains remote access through (e.g.) a misconfigured service;
(viii) {\em Trusted Relationship} (ATT\&CK T1199), whereby an attacker leverages the established trust relationship between modules to gain access;
(ix) {\em Malicious Commanding} (SPARTA IA-0007.02), whereby an attacker transmits a satellite command through a valid ground station; and
(x) {\em Exploit Code Flaw: Known Vulnerability} (SPARTA EX-0009.03), whereby an attacker exploits an unpatched software.

To determine $\beta(v,\at)$ and $\beta(e,\at)$, we leverage our domain knowledge in the same fashion as the SPARTA NRS~\cite{sparta_notional}.
Figure~\ref{fig:attackFlowGraph} highlights which $v\in V_{\general}$ and $e\in E_{\general}$ can be directly attacked by which $\at\in \AT_A$.
Likelihood $\beta(*,\at)$ is annotated with each $v$ and $e$, while noting $\beta(v,*)=0$ and $\beta(e,*)=0$ for all other $v$'s and $e$'s.
Specifically, we have
$\beta$(GM.A\&S, T1210) $=0.11$,
$\beta$(GM.CMD, T1199) $=0.33$,
$\beta$(GM.TX, T1199) $=0.11$,
$\beta$(GM.NET, T1595) $=0.40$,
$\beta$(GM.SOFT, T1592) $=0.11$,
$\beta$(GM.SOFT, T1566.001) $=0.07$,
$\beta$(SM.C\&DH, EX-0012) $=0.20$,
$\beta$(SM.C\&DH, EX-0009.03) $=0.17$,
$\beta$(SM.PAYCOM, IA-0007.02) $=0.45$,
$\beta$(GM.TX$\rightarrow$SM.BUSCOM, IA-0008.01) $=0.33$,
$\beta$(SM.BUSCOM$\rightarrow$GM.RX, REC-0005.02) $=0.33$,
$\beta$(SM.PAYCOM$\rightarrow$UM.RX, REC-0005.02) $=0.33$, and
$\beta$(SM.PAYCOM $\rightarrow$UM.RX, IA-0008.01) $=0.33$.
Note that the annotation $(0)$ on arc $e$  means $\beta(e,\at)=0$ for any $\at \in \AT_A$, meaning that $e$ cannot be directly attacked by any of the 10 techniques but is still relevant because $e$ can be compromised via cascading effects. Similarly, the annotation $(0)$ on node $v$ (e.g., UM.RX) means that $\beta(v,\at)=0$ for any $\at\in \AT_A$, namely that $v$ cannot be attacked by any of the 10 techniques but can be compromised via cascading effects.

In what follows we focus on repeating and mitigating Attack 1, while deferring Attacks 2 and 3 to
Appendix~\ref{appendix:Attacks-2-and-3}.

\subsubsection{Repeating Attack 1: Seizure of Control}
\label{sec:attack 3}

As highlighted by the dashed and dotted red arcs in Figure~\ref{fig:attackFlowGraph},
this attack employs 7 attack techniques and progresses as follows:
(i) the attacker employs  ATT\&CK technique T1592 to target the GM.SOFT module, where the attacker gains important information about the client computer (e.g., operating system version);
(ii) the attacker convinces the GM.SOFT user to execute a malicious file, such as malicious email
attachment (ATT\&CK T1566.001)
to entice the user;
(iii) the malware allows the attacker to scan (ATT\&CK T1595) GM.NET for additional hosts with open services;
(iv) the attacker uses the scan results to exploit a remote service (ATT\&CK T1210) in GM.A\&S, such as an overly permissive Samba service;
(v) the attacker laterally moves to GM.CMD by exploiting its trust relationship (ATT\&CK T1199) with GM.A\&S and prepositions the malware within a transmission request;
(vi) the attacker continues to laterally move to GM.TX by again exploiting its trust relationship (ATT\&CK T1199) with GM.CMD, embedding the malware in a command packet archive;
(vii) the attacker conducts a malicious command transmission (SPARTA IA-0007.02) from GM.TX to SM.PAYCOM to spread
the malware
to
the ISS; and
(viii) SM.PAYCOM pushes the command packet to the flight software bus where SM.C\&DH unpacks, parses, and executes the command, triggering the malware to exploit a known vulnerability (SPARTA EX-0009.03) at SM.C\&DH.
to gain control of this module.

We repeat this attack in our testbed as follows.
(i) We gain information about the operating system of the Software Access module GM.SOFT (ATT\&CK T1592) by enticing the space operator using the terminal to visit a malicious website.
(ii) We employ MSFvenom to embed a reverse shell in a space calculator utility to entice the space operator to download and execute our malware (ATT\&CK 1566.001).
(iii) With remote access to GM.SOFT, we employ Nmap to scan the Network Access module GM.NET through the VPN connection.
(iv) We discover that the Analysis \& Support module (GM.A\&S) runs a custom web server that is vulnerable to command injection and exploit this vulnerability (ATT\&CK T1210) to gain access to GM.A\&S.
(v) We design a beacon-based Command and Control (C2) bash script that piggybacks off the established Bus Management Mission Control Flow (Figure~\ref{fig:busMgmtGraph})
and the Telemetry Mission Data Flow (Figure~\ref{fig:telemGraph})
and embed it in a command to use the script execution function in the flight software
and place the command as a file in a directory on GM.A\&S, where
GM.CMD trusts (ATT\&CK T1199) and pulls files from GM.A\&S' shared directory to construct satellite commands.
(vi) Similarly,
GM.TX automatically receives the constructed command because it trusts GM.CMD (ATT\&CK T1199).
(vii) We accomplish our malicious command transmission (SPARTA IA-0007.02) because
SM.PAYCOM automatically receives the signal broadcast from GM.TX.
(viii) SM.PAYCOM pushes our command onto the flight software bus, where
SM.C\&DH retrieves and executes our command,
which exploits the flight software's lack of input validation (SPARTA EX-0009.03), leading to that our C2 script
implants itself in the SM.C\&DH module with root privileges. Our C2 script beacons back to GM.A\&S every time the satellite is within range of the Ground Station.

\subsection{Modeling and Analyzing Mission Risks}
\label{sec:cyber risk analysis}
We use Algorithms~\ref{alg:space cyber risks analysis} and~\ref{alg:mission disruption likelihood}
to assess the SATCOM mission disruption likelihood, with attack capabilities described in Section~\ref{sec:case study attacker capabilities},
$G_{\general}$  in Figure~\ref{fig:infrastructureGraph},
$G^{(i,1)}_{\mcf}$ for $1\leq i \leq m'_1=3$ and $G^{(i',1)}_{\mdf}$ for $1\leq i' \leq m'_1=2$ as described in Figures~\ref{fig:busMgmtGraph}-~\ref{fig:channelGraph}, respectively.
We run Algorithm~\ref{alg:space cyber risks analysis} on
an Alienware x15 R2 with an Intel i9-12900H processor and 32GB RAM.
We measure the execution time until the update on each element of the vector $(\L(v))_{v\in V_{\general}}$ is less than or equal to $ \epsilon=1^{-10}$.

In Case 0 ($\flag=0$), the nodes and arcs that cannot be attacked by any $\at \in \AT_A$ are susceptible to
cascading effects. In this case, Lines 16-17 are {\em not} executed and the results show that
$\L(v) > \tau$ for all 19 modules in $V_{\general}$
and $\L(e)\to 1$ for $e\in E_{\general}$.
This is expected because the lack of defense renders every vulnerable node and arc eventually compromised.
The measured execution time is 90 milliseconds, while noting that it only takes 42 iterations for the computation in the {\bf repeat} loop (Line 18) to converge.

In Case 1 ($\flag=1$), the nodes and arcs that cannot be attacked by any $\at \in \AT_A$ are not susceptible to cascading effects, meaning they can be disregarded in the subsequent space cyber risk analysis. In this case, Lines 16-17 are executed. The results show all
10 mission-required modules in the resulting $V_{\general}$ have $\L(v) > \tau$
and $\L(e)\to 1$ for the arcs in the resulting $E_{\general}$.
This is expected because the lack of defense renders every vulnerable node and arc eventually compromised. The measured execution time is 90 milliseconds, while it only takes 36 iterations for the computation in the {\bf repeat} loop (Line 18) to converge.
Note that for $e=(u,v)\in G_{\general}$, $\L(u)\leq \L(e)$ is possible because $e$, in addition to being affected by the compromise of $u$, is also subject to attacks directly targeting $e$ without compromising $u$.

To see the impact of cascading effects, we observe that in Case 0, there are 16 (out of the 19) modules with $\L(v)<\tau$ and 35 (out of the 36) arcs with $\L(e)< \tau$ {\em before} considering cascading effects (Lines 18-23); however, cascading effects result in $\L(v)>\tau$ for all 19 nodes
and $\L(e) \to 1$ for all 36 arcs.
In Case 1, there are 7 (out of the 10) nodes with $\L(v)<\tau$ and 13 (out of the 14) arcs with $\L(e)<\tau$ {\em before} considering  cascading effects (Lines 18-23); cascading effects result in $\L(v)>\tau$ for all 10 nodes
and $\L(e) \to 1$  for all 14 arcs.
This implies that
$A$ uses $G_{\general}$ to guide its attacks (i.e., effectively Case 0) causes more damage than using the $G_{\mcf}$'s and/or $G_{\mdf}$'s to guide its attacks,
because $G_{\general}$ offers more opportunities in leveraging cascading effects to compromise nodes and arcs with eventual impact on the $G_{\mcf}$'s and/or $G_{\mdf}$'s.

\begin{insight}
\label{insight:cascading effect}
Cascading effects offer attacker flexibilities in disrupting missions,
especially when attackers leverage the entire infrastructure to wage attacks (in contrast to only leveraging
the target control flows and data flows).
\end{insight}

Insight~\ref{insight:cascading effect}
means the defender should secure the modules that are not part of any mission control / data flows because attackers can pivot from them to attack the
targeted missions.

\subsection{Hardening Space Missions}

To demonstrate the actionability and usefulness of SCRM, now we show that if the defender is informed that the attacker may possess the 10 attack techniques mentioned above, then SCRM can identify and mitigate some (but not all) of the 10 attack techniques to cost-effectively thwart Attacks 1-3.

\begin{figure}[!htbp]
\centering{\includegraphics[width=.9\columnwidth]{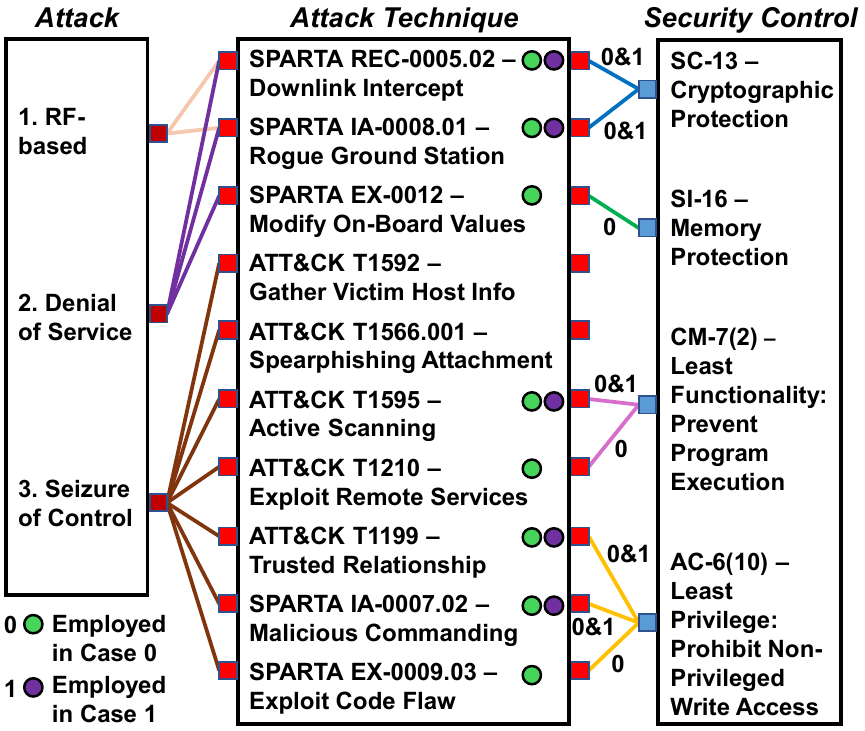}}
\caption{The 3 attacks, the 10 attack techniques they use, and the 4 security controls Algorithm~\ref{alg:mission hardening} identifies.}
\label{fig:mitigation}
\end{figure}

As discussed above,
Attacks 1-3
use 10 attack techniques
to disrupt $G^{(1,1)}_{\mcf}$, $G^{(2,1)}_{\mcf}$, $G^{(3,1)}_{\mcf}$, $G^{(1,1)}_{\mdf}$ and $G^{(2,1)}_{\mdf}$.
Now we show Algorithm~\ref{alg:mission hardening}
can identify and mitigate some of the 10 techniques (one input to SCRM) to
render the mission disruption likelihood to below $\tau=0.1$, effectively thwarting the three real-world attacks that are not known to the defender (i.e., the defender is only informed with the attack techniques, not specific attacks).
As summarized in Figure~\ref{fig:mitigation}, in Case 0, the
algorithm identifies and mitigate 8 (out of 10)
techniques
to reduce the mission disruption likelihood to $\L(j) = 0.04<\tau=0.1$; in Case 1, the algorithm identifies and mitigates 5 (out of 10) techniques
to reduce the mission disruption likelihood $\L(j) = 0.08<\tau=0.1$. The 5 techniques in Case 1 are a subset of the 8 in Case 0 but it is an open problem to (dis)prove that this is universally true.
As shown in Figure~\ref{fig:mitigation}, it suffices to (manually) identify 4 security controls to mitigate the 8 techniques and
render $\L(j=1)<\tau$.

Now we describe how the mitigation thwarts Attack 1, while deferring the details of our experiments for mitigating Attacks 2 and 3 to Appendix~\ref{appendix:Attacks-2-and-3}.
Our experiment for mitigating Attack 1 (Seizure of Control) proceeds as follows.
In Case 0, we employ CM-7(2) (Least Functionality: Prevent Program Execution) via an application denylist to inspect operating system processes and file writes at the GM.NET, GM.A\&S, GM.CMD, GM.TX, SM.PAYCOM, and SM.C\&DH modules.
Our attempt to execute T1595 (using Nmap to conduct active scanning) at GM.NET failed because of the mitigation.
Our attempt to execute T1210 (spawning Python and Bash terminals through a command injection vulnerability) at GM.A\&S is instantly blocked by the security control.
We also employ AC-6(10) (Least Privilege: Prohibit Non-Privileged Write Access) at GM.CMD, GM.TX, SM.PAYCOM, and SM.C\&DH with immutable file settings to prevent command files from tampering.
Our attempt to execute T1199 at GM.CMD and GM.TX (manipulating their data storage with a malicious satellite command file) failed because AC-6(10) blocks the file write operation.
Our attempt to locally execute IA-0007.02 (using a python script to inject a malicious flight command) at SM.PAYCOM failed because of the mitigation.
Our attempt to execute EX-0009.03 (executing a malicious flight command) at SM.C\&DH via a python script to place the malicious command on the software bus also failed because of AC-6(10).
That is, Attack 1 is largely mitigated.
In Case 1, the same experiments and results apply except that we only need to employ security control CM-7(2) at GM.NET to prevent T1595, and employ security control AC-6(10) at GM.CMD to prevent T1199 and at SM.PAYCOM to prevent IA-0007.02, because they close all known routes for attackers to gain access from GM.SOFT to
GM.A\&S, GM.TX, and SM.C\&DH.

As highlighted in Figure~\ref{fig:mitigation}, in Case 0, two (2) attack techniques (T1592 and T1566.001) are not mitigated, explaining the {\em residual} risk of
$\L(j) = 0.04$;
in Case 1, five (5) techniques (EX-0012,
T1592, T1566.001,
T1210,
and EX-0009.03)
are not mitigated,
leading to residual risk of $\L(j) = 0.08$.

\begin{insight}
\label{insight:hardening}
Our framework can effectively harden missions by mitigating {\em some}, but not necessarily all, attack techniques that are possibly possessed by the attacker. Moreover, attack cascading effects demand the mitigation of more attack techniques (e.g., $8-5=3$ in our case study).
\end{insight}

Insight~\ref{insight:hardening} is important because it is significant to mitigate fewer than $|\AT_A|$ attack techniques (i.e., incurring a lower cost). This cost-effectiveness is amplified when CTI does not provide an accurate $|\AT_A|$, meaning defender is forced to consider a set $|\AT_A|$ that is much larger than necessary.

\subsection{Comparison of SCRM with NRS}\label{subsec:nrs-comparison}

To show SCRM substantially advances state of the art
space cyber risk management solutions (i.e., NRS), we show how SCRM accomplishes what NRS cannot do. For this purpose, we use the Seizure of Control attack described in Section \ref{sec:attack 3} against the SATCOM infrastructure described in Section~\ref{subsec:TestbedInfrastructure}.

\noindent{\bf NRS-based risk analysis and mission hardening of the scenario.}
NRS assigns a scalar risk score to each  SPARTA attack technique in isolation~\cite{sparta_notional}. That is, NRS would independently score techniques such as Malicious Commanding (IA-0007.02) against the Payload Communications module (SM.PAYCOM) and Exploit Code Flaw (EX-0009.03) against the Command \& Data Handling module (SM.C\&DH). This process requires a subject-matter expert at four of its six steps~\cite{Ear2024NRS, XuIEEECSR2025-NRS}. More importantly, NRS treats each module in isolation: the score for SM.C\&DH reflects only the direct exploitability of that module, with no mechanism to propagate the consequence of first compromising GM.SOFT, then GM.NET, then GM.AS, and so on through the chain that ultimately reaches SM.C\&DH; such cascading risks are not discussed in \cite{Ear2024NRS, XuIEEECSR2025-NRS}. This is because NRS does not model the communication graph of the infrastructure.
The consequence is that NRS {\em systematically underestimates} risk for attack paths that span multiple segments and underestimates the scope of the defender's required response. Because NRS scores techniques independently against individual modules, a defender relying on NRS would not be able to determine whether addressing certain technique module pairs would reduce mission risk to an acceptable level. As such, NRS would recommend security controls for all techniques.

\smallskip

\noindent{\bf SCRM-based risk analysis and mission hardening of the scenario.} As described in Section \ref{sec:cyber risk analysis},
SCRM embeds the same set of attack techniques into the directed multigraph $G_{\general}$ of the full SATCOM infrastructure ($|V_{\general}| = 19$ modules, $|E_{\general}| = 36$ arcs). Algorithm~\ref{alg:space cyber risks analysis} first computes the direct compromise likelihood of each node and arc from the technique possession likelihoods and the direct compromise likelihoods. It then propagates cascading effects (Types 1-3, Section~\ref{sec:cascading-effects}) until convergence. In Case 0 (flag $= 0$), 3 of the 19 modules have risk greater than the set risk threshold before cascading effects are considered. After convergence, all 19 modules exceed the threshold. We conclude that this is due to the lack of defense which renders every vulnerable node and arc eventually compromised via a trust chain between them. Importantly, an NRS evaluation, which only considers direct effects, would conclude that 16 modules are adequately protected, when cascading propagation leaves none of them safe. SCRM bridges this gap without additional expert judgment.
SCRM, in contrast to NRS, ties hardening decisions directly to mission disruption likelihood via Algorithm~\ref{alg:mission hardening}. In the case study, Algorithm~\ref{alg:mission hardening} determines that mitigating 8 of the 10 techniques reduces the risk below the mission threshold. Additionally it identifies which NIST SP 800-53 security controls should be employed at each relevant node and arc, while noting that the remaining risk is attributable to the 2 unmitigated techniques. An NRS-guided process provides no equivalent because NRS lacks a mission model and therefore cannot determine whether any combination of countermeasures is sufficient to protect a mission's control and data flows.

Summarizing the preceding discussion, we draw:

\begin{insight}
The state of the art NRS-based space cyber risks analysis would substantially underestimate the risk (missing 16 modules identified as risk inducing through cascading effects), while substantially overestimate the investment in hardening missions (recommended mitigations for all 10 techniques instead of the required 8 to reduce risk below the mission threshold).
\end{insight}

\section{Limitations and Future Directions}\label{sec:limitations}

This study
has 7 limitations. {\bf (i)} SCRM currently focuses on dealing with likelihood of compromises without considering impact of compromises.
It is important to extend the study to accommodate non-constant impacts.
{\bf (ii)} Algorithm~\ref{alg:mission hardening} is by no means optimal, meaning that it may mitigate more attack techniques than necessary. It is important to design optimal hardening algorithms.
{\bf (iii)} When dealing with aggregation of attack effects, we make simplifying assumptions via the use of the $\max$ function. This is shown by Eq.\eqref{eq:L(j) aggregation function}, which aggregates attack effects over different control flows and data flows, and by Lines 2 and 4 in Algorithm~\ref{alg:mission disruption likelihood}, which aggregate attack effects over the nodes and arcs in a control or data flow. Such choices of aggregation function need experimental validation.
{\bf (iv)} When dealing with cascading effects of successful attacks, we also make simplifying assumptions, especially the independence assumptions behind Eqs.\eqref{eq:function-f}-\eqref{eq:function-h'}. It is an outstanding and challenging open problem to cope with the dependence that might exist, because there are different kinds of dependence structures.
{\bf (v)} Our case study focuses on $n=1$ mission, despite it has $m'_1=3$ control flows and $n'_1=2$ data flows. This is imposed by the scale of our testbed, which we plan to expand, permitting us to investigate more mission scenarios. Moreover, we manually create the concrete model, including $G_{\general}$, $G_{\mcf}$, and $G_{\mdf}$, based on our domain knowledge. It is an outstanding open problem to automatically extract these graph structures either from their specifications and/or code.
{\bf (vi)} Our case study considers 2 extreme cases: attack cascading effects apply to all (Case 0) vs. none (Case 1) of the modules that are not directly affected by the attack techniques possibly possessed by attackers. There could be scenarios in between.
{\bf (vii)} We focus on  mission disruption incurred by space cyber attacks. Including additional sources of risks (e.g., space weather) would make SCRM more widely applicable.

\section{Conclusion}
\label{sec:conclusion}

We
presented a novel framework for  space cyber risk management, including a formal model of space infrastructures, their missions, and cyber attacks against them.
We conduct experiments on our space testbed to demonstrate the framework's usefulness, by re-implementing 3 real-world space cyber attacks, using our framework (more specifically its mission hardening algorithm) to identify security countermeasures, and employing our countermeasures to effectively mitigate the space cyber risks. Future research problems are abundant, including the limitations of the present study. We hope that this study will inspire many more efforts from the research community to tackle the space cybersecurity problem.

\cleardoublepage

\appendix

\subsection{Repeating and Mitigating Attacks 2-3}
\label{appendix:Attacks-2-and-3}

\subsubsection{Attack 2: DoS Attack}
\label{sec:attack 2}

As highlighted by the dotted blue arcs in Figure~\ref{fig:attackFlowGraph},
this attack employs 3 attack techniques as follows:
(i) the attacker targets the arc SM.BUSCOM$\to$GM.RX, by intercepting
the legitimate data signals (SPARTA REC-0005.02) from the Galaxy 4 satellite;
(ii) the attacker uses its rogue ground station (SPARTA IA-0008.01) to send a malicious command to SM.BUSCOM; and
(iii) SM.BUSCOM automatically forwards the malicious command to SM.C\&DH, which parses the command
and executes it, corrupting its on-board configuration (SPARTA
EX-0012) and
halting
the flight computer.

We repeat the DoS Attack
in our testbed as follows.
(i) We intercept the downlink (SPARTA REC-0005.02) at SM.BUSCOM$\to$GM.RX.
(ii) We craft a malicious flight command to trigger a memory access violation in the flight software and use
a HackRF as a rogue ground station (SPARTA IA-0008.01) to transmit the signal to SM.BUSCOM.
(iii) SM.BUSCOM automatically places the malicious command on the flight software bus, where SM.C\&DH retrieves, parses, and executes the malicious command (SPARTA EX-0012) to cause (e.g.) a buffer overflow, crashing the flight software.

\noindent{\bf Mitigation of Attack 2}. Our experiment for showing how the mitigation largely thwarts Attack 2 (DoS) proceeds as follows.
In Case 0, we use security control SC-13 to protect arc SM.BUSCOM$\to$ GM.RX against data eavesdropping technique REC-0005.02  and arc GM.TX$\to$SM.BUSCOM against data injection technique IA-0008.01. Our attempt to eavesdrop the communication over
SM.BUSCOM$\to$GM.RX failed because we cannot decrypt the demodulated signal without having the key. Our attempt to inject commands over arc SM.BUSCOM$\to$GM.RX also failed because the injected data cannot pass the message authentication verification.
We also employ security control SI-16 (Memory Protection) at the SM.C\&DH module to run an on-orbit intrusion protection system (IPS) via allowlists and denylists
(e.g., allowing commands that conform to the CCSDS Telecommand protocol, while denying commands that attempt to access the underlying operating system).
Our attempt to apply technique EX-0012 (triggering the buffer overflow) at SM.C\&DH failed because security control SI-16 detects and blocks the attempt.
Thus, Attack 2 is completely thwarted.
In Case 1, we only need to employ security control SC-13 to protect SM.BUSCOM$\to$GM.RX against REC-0005.02 and protect GM.TX$\to$SM.BUSCOM against IA-0008.01 (i.e., we do not need to employ SI-16
at SM.C\&DH because the employed SC-13
has already closed all known routes to gaining access to SM.C\&DH.

\subsubsection{Attack 3: RF-based Attack}
\label{sec:attack 1}

As highlighted by the dashed brown arcs in Figure~\ref{fig:attackFlowGraph},
this attack employs 2 attack techniques (i.e., REC-0005.02 and IA-0008.01)
and progresses as follows:
(i) the attacker targets the arc SM.PAYCOM$\to$UM.RX, where the attacker
intercepts legitimate downlink broadcasts (SPARTA REC-0005.02) of an Israeli satellite TV channel to identify the properties of the legitimate data signal; and
(ii) the attacker targets the same arc
to generate the attacker's own broadcast content
via its rogue ground station (IA-0008.01) to overpower the legitimate signal.

We repeat this attack
in our testbed as follows.
(i) We equip our attacking terminal with a HackRF, which is programmed to scan through Multi-Use Radio Service (MURS) frequencies, and position the terminal in the vicinity of the legitimate user terminal to intercept downlink signals (SPARTA REC-0005.02); upon receipt of a valid signal, we analyze the signal to understand the modulation, data packing, and transfer protocols used, while assuming the communications data is not encrypted (which is not uncommon in the space domain~\cite{yu2024comprehensive,willbold2023space}).
(ii) We construct a competing signal and broadcast it (SPARTA IA-0008.01) at a gain level appropriate to overpower the legitimate satellite (SM.PAYCOM), while not over-saturating the victim user terminal (UM.RX).

\noindent{\bf Mitigation of Attack 3}. Our experiment for showing how the mitigation thwarts Attack 1, namely the RF-based Attack, proceeds as follows.
In Case 0, we implement
security control SC-13 (Cryptographic Protection), a python-based implementation of
AES-256 together with HMAC/SHA-256, to mitigate technique REC-0005.02 and protect the arc SM.PAYCOM$\to$UM.RX from both eavesdropping and manipulation attacks. After employing this mitigation, we re-wage Attack 1 to intercept and inspect the signal transmitted over arc SM.PAYCOM$\to$UM.RX, only finding that the signal is cryptographically protected, meaning REC-0005.02 is thwarted.
This mitigation also thwarts IA-0008.01 (injecting malicious data / signal over arc SM.PAYCOM$\to$UM.RX), as shown in our experiment that our injection attempt fails because the injected data cannot pass the message authentication verification.
In Case 1, the same experiment and result apply when using security control SC-13 to mitigate  REC-0005.02 and IA-0008.01.

\subsection{List of Acronyms, Abbreviations, and Identifications}
\label{appendix:acronyms}
\begin{tabular}{>{\raggedright\arraybackslash}p{2cm} >{\raggedright\arraybackslash}p{5cm}}
A\&S & Analysis \& Support (Module) \\
ATCTRL & Attitude Control (Module) \\
ATT\&CK & Adversarial Tactics, Techniques, and Common \\&Knowledge \\
BUSCOM & Bus Communications (Module) \\
C\&DH & Command \& Data Handling (Module) \\
Casc. & Cascading Effects \\
CMD & Command (Module) \\
CSF & Cyber Security Framework \\
EX-0009.03 & Exploit Code Flaw: Known Vulnerability \\
EX-0012 & Modify On-Board Values (SPARTA) \\
GM & Ground Module \\
IA-0007.02 & Malicious Commanding (SPARTA) \\
IA-0008.01 & Rogue Ground Station (SPARTA) \\
MA & Mission Analysis (Module) \\
NET & Network Access (Module) \\
NRS & Notional Risk Score \\
PAYCOM & Payload Communications (Module) \\
PAYPROC & Ground Payload Processing (Module) \\
PNT & Positioning, Navigation and Timing \\
POWER & Electrical Power (Module) \\
PROC & Processing (Module) \\
PROP & Propulsion (Module) \\
REC-0005.02 & Downlink Intercept (SPARTA) \\
RX & Reception (Module) \\
SCRM & Space Cyber Risk Managment \\
SM & Space Module \\
SOFT & Software Access (Module) \\
SPARTA & Space Attack Research and Tactic Analysis \\
T1199 & Trusted Relationship (ATT\&CK) \\
T1210 & Exploitation of Remote Services (ATT\&CK) \\
T1566.001 & Spearphishing Attachment (ATT\&CK) \\
T1592 & Gather Victim Host Information (ATT\&CK) \\
T1595 & Active Scanning (ATT\&CK) \\
TELEM & Telemetry (Module) \\
THCTRL & Thermal Control (Module) \\
TX & Transmission (Module) \\
UM & User Module
\end{tabular}

\newpage 
\bibliographystyle{plain}
\bibliography{references}

@article{Ear2024NRS,
  author       = {Ekzhin Ear and Brandon Bailey and Shouhuai Xu},
  title        = {Towards Principled Risk Scores for Space Cyber Risk Management},
  journal      = {CoRR},
  volume       = {abs/2402.02635},
  year         = {2024},
}

@online{sparta_homepage,
  author = {{Aerospace Corporation}},
  title = {Space Attack Research \& Tactic Analysis {(SPARTA)}},
  year = {2023},
  url = {https://sparta.aerospace.org},
  urldate = {2023-12-19}
}

@online{sparta_notional,
  author = {{Aerospace Corporation}},
  title = {Notional Risk Scores},
  year = {2023},
  url = {https://sparta.aerospace.org/notional-risk-scores},
  urldate = {2023-12-19}
}

@article{fritz2013satellite,
  title={Satellite hacking: A guide for the perplexed},
  author={Fritz, Jason},
  journal={Culture Mandala},
  volume={10},
  number={1},
  pages={5906},
  year={2013},
  publisher={Bond University}
}

@incollection{jones2022risk,
  title={A Risk Based Approach to Space System Protection},
  author={Jones, Michael P and Gilbert, Kevin W},
  booktitle={ASCEND 2022},
  pages={4300},
  year={2022}
}

@manual{NASA-STD-1006A,
  title        = "{NASA-STD-1006A}: Space System Protection Standard",
  author       = "{NASA}",
  organization = "{NASA}",
  address      = "Hampton, VA",
  year         = 2022
}

@manual{NASA-SS-BPG,
  title        = "{Space Security: Best Practices Guide}",
  author       = "{NASA}",
  organization = "{NASA}",
  address      = "Hampton, VA",
  year         = 2023
}

@incollection{falco2021security,
  title={A security risk taxonomy for commercial space missions},
  author={Falco, Gregory and Boschetti, Nicolo},
  booktitle={ASCEND 2021},
  pages={4241},
  year={2021}
}

@inproceedings{knez2016lessons,
  title={Lessons learned from applying cyber risk management and survivability concepts to a space mission},
  author={Knez, Claudia and Llans{\'o}, Thomas and Pearson, Dallas and Schonfeld, Tibor and Sotzen, Kristin},
  booktitle={2016 IEEE Aerospace Conference},
  pages={1--8},
  year={2016},
  organization={IEEE}
}

@article{bailey2021cybersecurity,
  title={Cybersecurity Protections for Spacecraft: A Threat Based Approach},
  author={Bailey, Brandon},
  journal={Aerospace Corporation},
  year={2021}
}

@misc{joint2012nist800-30,
  title={{NIST SP 800-30r1:} Guide for Conducting Risk Assessments},
  author={{Joint Task Force Transformation Initiative}},
  year={2012},
  publisher={September}
}

@misc{joint2018nist800-37,
  title={{NIST SP 800-37r2:} Risk Management Framework for Information Systems and Organizations: A System Life Cycle Approach for Security and Privacy},
  author={{Joint Task Force}},
  year={2018},
  publisher={December}
}

@inproceedings{ear2023characterizing,
  title={Characterizing Cyber Attacks against Space Systems with Missing Data: Framework and Case Study},
  author={Ear, Ekzhin and Remy, Jose LC and Feffer, Antonia and Xu, Shouhuai},
  booktitle={2023 IEEE Conference on Communications and Network Security (CNS)},
  pages={1--9},
  year={2023},
  organization={IEEE}
}

@article{mccarthy2023cybersecurity,
  title={Cybersecurity Framework Profile for Hybrid Satellite Networks (HSN)},
  author={McCarthy, James and Mamula, Dan and Brule, Joseph and Meldorf, Karri and Jennings, Rory and Wiltberger, John and Thorpe, Chris and Dombrowski, John and Lattin, O'Ryan and Sepassi, Sam and others},
  journal={National Institute of Standards and Technology, NIST Interagency or Internal Report (IR) NIST IR},
  volume={8441},
  number={2023},
  pages={28},
  year={2023}
}

@article{scholl2021introduction,
  title={Introduction to cybersecurity for commercial satellite operations},
  author={Scholl, Matthew and Suloway, Theresa},
  journal={National Institute of Standards and Technology, Tech. Rep},
  year={2021}
}

@article{bartock2021foundational,
  title={Foundational PNT profile: Applying the cybersecurity framework for the responsible use of positioning, navigation, and timing (PNT) services},
  author={Bartock, Michael and Brule, Joseph and Li-Baboud, Y and Lightman, Suzanne and McCarthy, James and Reczek, Karen and Northrip, Doug and Scholz, Arthur and Suloway, Theresa},
  journal={National Institute of Standards and Technology (NIST)},
  year={2021}
}

@techreport{lightman2022satellite,
  title={Satellite Ground Segment: Applying the Cybersecurity Framework to Assure Satellite Command and Control},
  author={Lightman, Suzanne and Suloway, Theresa and Brule, Joseph},
  year={2022},
  institution={National Institute of Standards and Technology}
}

@article{NIST800-53A,
  title={Assessing security and privacy controls in information systems and organizations},
  author={Force, Joint Task},
  journal={NIST Special Publication},
  volume={800},
  pages={53A},
  year={2022}
}

@techreport{NIST800-53r5,
  title={Security and privacy controls for information systems and organizations},
  author={Force, Joint Task},
  journal={NIST Special Publication},
  volume={800},
  pages={53},
  year={2020},
  institution={National Institute of Standards and Technology}
}

@article{NIST800-53B,
  title={Control baselines for information systems and organizations},
  author={Force, Joint Task},
  journal={NIST Special Publication},
  volume={800},
  pages={53B},
  year={2020}
}

@manual{NISTCSF2,
  title        = "{The {NIST} Cybersecurity Framework, Version 2.0}",
  author       = "{National Institute of Standards and Technology}",
  organization = "{National Institute of Standards and Technology}",
  address      = "Gaithersburg, MD",
  year         = "2024",
  month        = "February",
  url          = "https://doi.org/10.6028/NIST.CSWP.29"
}

@article{pavur2022building,
  title={Building a launchpad for satellite cyber-security research: lessons from 60 years of spaceflight},
  author={Pavur, James and Martinovic, Ivan},
  journal={Journal of Cybersecurity},
  volume={8},
  number={1},
  pages={tyac008},
  year={2022},
  publisher={Oxford University Press}
}

@article{guo2021survey,
  title={A survey on space-air-ground-sea integrated network security in 6G},
  author={Guo, Hongzhi and Li, Jingyi and Liu, Jiajia and Tian, Na and Kato, Nei},
  journal={IEEE Communications Surveys \& Tutorials},
  volume={24},
  number={1},
  pages={53--87},
  year={2021},
  publisher={IEEE}
}

@article{tedeschi2022satellite,
  title={Satellite-based communications security: A survey of threats, solutions, and research challenges},
  author={Tedeschi, Pietro and Sciancalepore, Savio and Di Pietro, Roberto},
  journal={Computer Networks},
  pages={109246},
  year={2022},
  publisher={Elsevier}
}

@inproceedings{willbold2023space,
  title={Space odyssey: An experimental software security analysis of satellites},
  author={Willbold, Johannes and Schloegel, Moritz and V{\"o}gele, Manuel and Gerhardt, Maximilian and Holz, Thorsten and Abbasi, Ali},
  booktitle={2023 IEEE Symposium on Security and Privacy (SP)},
  pages={1--19},
  year={2023},
  organization={IEEE}
}

@inproceedings{yu2024comprehensive,
  title={A Comprehensive Analysis of Security Vulnerabilities and Attacks in Satellite Modems},
  author={Yu, Lingjing and Hao, Jingli and Ma, Jun and Sun, Yong and Zhao, Yijun and Luo, Bo},
  booktitle={Proceedings of the 2024 on ACM SIGSAC Conference on Computer and Communications Security},
  pages={3287--3301},
  year={2024}
}

@inproceedings{falco2023wannafly,
  title={Wannafly: An approach to satellite ransomware},
  author={Falco, Gregory and Thummala, Rajiv and Kubadia, Arpit},
  booktitle={2023 IEEE 9th International Conference on Space Mission Challenges for Information Technology (SMC-IT)},
  pages={84--93},
  year={2023},
  organization={IEEE}
}

@inproceedings{hansen2024guarding,
  title={Guarding the Galaxy: Satellite Ransomware and Countermeasures},
  author={Hansen, Petersen and Henry, Wayne C and Reith, Mark G and Thummala, Rajiv and Falco, Gregory},
  booktitle={2024 IEEE Aerospace Conference},
  pages={1--6},
  year={2024},
  organization={IEEE}
}

@inproceedings{falco2021cubesat,
  title={Cubesat security attack tree analysis},
  author={Falco, Gregory and Viswanathan, Arun and Santangelo, Andrew},
  booktitle={2021 IEEE 8th International Conference on Space Mission Challenges for Information Technology (SMC-IT)},
  pages={68--76},
  year={2021},
  organization={IEEE}
}

@inproceedings{boschetti2023ground,
  title={Ground station as a service reference architectures and cyber security attack tree analysis},
  author={Boschetti, Nicol{\`o} and Smethurst, Chelsea and Epiphaniou, Gregory and Maple, Carsten and Sigholm, Johan and Falco, Gregory},
  booktitle={2023 IEEE Aerospace Conference},
  pages={1--12},
  year={2023},
  organization={IEEE}
}

@incollection{falco2020satellites,
  title={When satellites attack: Satellite-to-satellite cyber attack, defense and resilience},
  author={Falco, Gregory},
  booktitle={ASCEND 2020},
  pages={4014},
  year={2020}
}

@inproceedings{falco2023scrub,
  title={How to scrub a launch: spaceport cybersecurity},
  author={Falco, Gregory and Korth, Luke and Custer, Patrick and Schofield, Rembert N and Pocock, Caleb},
  booktitle={2023 IEEE 9th International Conference on Space Mission Challenges for Information Technology (SMC-IT)},
  pages={56--67},
  year={2023},
  organization={IEEE}
}

@inproceedings{lee2023vulnerabilities,
  title={The Vulnerabilities Less Exploited: Cyberattacks on End-of-Life Satellites.},
  author={Lee, Frank and Falco, Gregory},
  booktitle={SpaceSec},
  year={2023}
}

@incollection{ramdass2022cybersecurity,
  title={Cybersecurity of On-Orbit Servicing, Assembly, and Manufacturing (OSAM) Systems},
  author={Ramdass, Tyler J and Munshi, Ninad and Kim, Richard and Falco, Gregory},
  booktitle={ASCEND 2022},
  pages={4379},
  year={2022}
}

@INPROCEEDINGS{XuIEEESP2025,
  author={Remy, Jose Luis Castanon and Ear, Ekzhin and Chang, Caleb and Feffer, Antonia and Xu, Shouhuai},
  booktitle={2025 IEEE Symposium on Security and Privacy (SP)}, 
  title={SoK: Space Infrastructures Vulnerabilities, Attacks and Defenses}, 
  year={2025},
  volume={},
  number={},
  pages={1028-1046},
  doi={10.1109/SP61157.2025.00138}}

@MISC{ATTCK,
AUTHOR="Blake Strom",
TITLE="ATT\&CK 101: Cyber Threat Intelligence",
YEAR="2018",
URL="{https://www.mitre.org/capabilities/cybersecurity/overview/cybersecurity-blog/attck-101}"
}

@article{XuIEEEACMToN2019,
  author    = {Z. Lin and
               W. Lu and
               S. Xu},
  title     = {Unified Preventive and Reactive Cyber Defense Dynamics Is Still Globally Convergent},
  journal   = {IEEE/ACM ToN},
  volume    = {27},
  number    = {3},
  pages     = {1098--1111},
  year      = {2019},
}

@inproceedings{XuIEEECSR2025-NRS,
  author       = {Ekzhin Ear and Brandon Bailey and Shouhuai Xu},
  title        = {The Notional Risk Scores Approach to Space Cyber Risk Management},
  booktitle    = {IEEE Conference on Cyber Security and Resilience (CSR'2025)},
  year         = {2025},
}

@article{XuIEEETNSE2018,
  author    = {R. Zheng and
               W. Lu and
               S. Xu},
  title     = {Preventive and Reactive Cyber Defense Dynamics Is Globally Stable},
  journal   = {IEEE TNSE},
  volume    = {5},
  number    = {2},
  pages     = {156--170},
  year      = {2018},
}

@article{XuInternetMath2015Dependence,
  author    = {M. Xu and G. Da and S. Xu},
  title     = {Cyber Epidemic Models with Dependences},
  journal   = {Internet Mathematics},
  volume    = {11},
  number    = {1},
  pages     = {62--92},
  year      = {2015},
}

@inproceedings{XuSpaceSec2025,
  author       = {Jose Luis Castanon Remy and Caleb Chang and Ekzhin Ear and Shouhuai Xu},
  title        = {Space Cybersecurity Testbed: Fidelity Framework, Example Implementation, and Characterization},
  booktitle    = {Proc. of SpaceSec},
  year         = {2025},
}

\end{document}